# The global atmospheric electrical circuit and climate

## R.G. Harrison

**Department of Meteorology, The University of Reading**
**P.O. Box 243, Earley Gate, Reading, Berkshire, RG6 6BB, UK**

*Email*: r.g.harrison@reading.ac.uk

**Abstract**
Evidence is emerging for physical links among clouds, global temperatures, the global atmospheric electrical circuit and cosmic ray ionisation. The global circuit extends throughout the atmosphere from the planetary surface to the lower layers of the ionosphere. Cosmic rays are the principal source of atmospheric ions away from the continental boundary layer: the ions formed permit a vertical conduction current to flow in the fair weather part of the global circuit. Through the (inverse) solar modulation of cosmic rays, the resulting columnar ionisation changes may allow the global circuit to convey a solar influence to meteorological phenomena of the lower atmosphere. Electrical effects on non-thunderstorm clouds have been proposed to occur via the ion-assisted formation of ultrafine aerosol, which can grow to sizes able to act as cloud condensation nuclei, or through the increased ice nucleation capability of charged aerosols. Even small atmospheric electrical modulations on the aerosol size distribution can affect cloud properties and modify the radiative balance of the atmosphere, through changes communicated globally by the atmospheric electrical circuit. Despite a long history of work in related areas of geophysics, the direct and inverse relationships between the global circuit and global climate remain largely quantitatively unexplored. From reviewing atmospheric electrical measurements made over two centuries and possible paleoclimate proxies, global atmospheric electrical circuit variability should be expected on many timescales.

**Summary**

The atmospheric electrical circuit is modulated by variations in shower clouds and tropical thunderstorms, and generates a vertical electric field in non-thunderstorm (or *fair weather*) regions globally. Modulation of the fair weather electric field occurs on daily, seasonal, solar cycle and century timescales. There is a long history of atmospheric electric field measurements.

Variations in the atmospheric electrical circuit have a new relevance because of empirical relationships found between cosmic ray ion production, cloud properties and global temperature. Mechanisms have been proposed linking clouds and the solar modulation of cosmic rays through the microphysics of ions, aerosols and clouds. Theories exist which explain ion-assisted aerosol particle formation and the increased capture rates of charged aerosols over neutral aerosols by cloud droplets, possibly enhancing droplet freezing. Atmospheric electrical modification of cloud properties may have significant global implications for climate, via changes in the atmospheric energy balance.

Combining studies of cloud microphysics and cosmic rays with the global atmospheric electrical circuit presents a relatively unexplored area of climate science, in which progress can clearly be expected. Events such as the atmospheric nuclear weapons tests offer, with suitable theory and cloud modelling, the possibility to quantify changes in radiative properties of the atmosphere associated with atmospheric electrical changes. The long-term condition of the atmospheric electrical system may be inferred from a combination of historical measurements and proxies, and compared with known variations in global temperature.

# 1 Introduction

## *1.1 Scope and motivation*

Atmospheric electricity is one of the longest-investigated geophysical topics, with a variety of measurement technologies emerging in the late eighteenth century and reliable data available from the nineteenth century. Of modern relevance, however, is the relationship of atmospheric electricity to the climate system (Bering, 1995), through climate-induced changes to thunderstorms (Williams 1992, 1994) and the electrical modification of non-thunderstorm clouds. Although it is well-established that clouds and aerosol modify the local atmospheric electrical parameters (Sagalyn and Faucher, 1954), aerosol microphysics simulations (Yu and Turco, 2001) and analyses of satellite-derived cloud data (Marsh and Svensmark, 2000) now suggest that aerosol formation, coagulation and in-cloud aerosol removal could themselves be influenced by changes in the electrical properties of the atmosphere (Harrison and Carslaw, 2003). Simulations of the twentieth century climate underestimate the observed climate response to solar forcing (Stott *et al*., 2003), for which one possible explanation is a solar-modulated change in the atmospheric electrical potential gradient (PG) affecting clouds and therefore the radiative balance. As with many atmospheric relationships, establishing cause and effect from observations is complicated by the substantial natural variability present. The importance of assessing the role of solar variability in climate makes it timely to review what is known about the possible relevance of the atmospheric electrical circuit of the planet to its climate. The scope of this paper therefore includes the physical mechanisms by which global atmospheric electricity may influence aerosols or clouds, and ultimately climate. This is *not* a review of thunderstorm electrification, but a discussion of possible influences on the global atmospheric electrical circuit, and the atmospheric and climate processes it may influence in turn. The considerable sensitivity of the planet's albedo to cloud droplet concentrations (Twomey, 1974), presents a strong motivation for investigating possible electrical effects on cloud microphysics.

Several, traditionally distinct, geophysical topics have to be considered together in order to make progress in the interdisciplinary subject area of solar-terrestrial physics, atmospheric electricity and climate. First, the atmospheric electrical circuit has to be understood, as it communicates electrical changes globally throughout the weather-forming regions of the troposphere. Secondly, changes in thunderstorms and shower clouds caused by surface temperature changes are likely to provide an important modulation on the global atmospheric electrical circuit. Thirdly, the microphysics of clouds, particularly ice nucleation and water droplet formation on aerosol particles has to be assessed in terms of which mechanisms, in a myriad of other competing and complicated cloud processes, are the most likely to be significantly affected by electrical changes in the atmosphere. Changes in the global properties of clouds, even to a small extent, have implications for the long-term energy balance of the climate system: electrically-induced cloud changes present a new aspect (Kirkby, 2001). Fourthly, galactic cosmic rays, which are modulated by solar activity, provide a major source of temporal and spatial variation in the atmosphere's electrical properties. The cosmic ray changes include sudden reductions and perturbations on the timescales of hours (Forbush decreases and solar proton events) as well as variability on solar cycle (~decadal) time scales and longer. The integration of these four disparate subject areas is a major geophysical challenge, but, as this paper shows, the elements exist for an integrated quantitative understanding of the possible connections between solar changes, cosmic ray ionisation, the global atmospheric electrical circuit and climate.

Figure 1 summarises the geophysical processes considered in this paper, which has been drawn to illustrate the links between the global atmospheric electrical circuit and the climate system. In developing these ideas, Section 2 summarises the global electrical system and Section 3 discusses the influences of cosmic rays on the atmosphere, knowledge of which may permit extending analysis of the atmospheric electrical system back in geological time. Fossil evidence of ancient atmospheric electrical activity on Earth (Harland and Hacker, 1966) indicates that electrification is not just a recent property of the terrestrial atmosphere: the possibility of solar modulation of climate through the global circuit therefore exists into the distant past.

In the remainder of Section 1, special attention is given to the pioneering early measurements made in the subject. Although different land stations rarely show correlations with each other on an hourly basis (Bhartendu, 1971), they form the majority of atmospheric electrical measurements available, motivating the development of new data processing methods (Harrison, 2004a, 2004b). Section 2 considers aspects of the global atmospheric electrical circuit, and its modulation on different timescales. Section 3 summarises the physics of the links between the global circuit and climate, and Section 4 suggest methods by which they can be investigated further.

### *1.2 Historical overview of Fair Weather atmospheric electricity*

Observations of a sustained, but unexplained electrification in the fair weather atmosphere probably provided the motivation for the early research. Subsequently, the sensitivity of electrical atmospheric parameters to meteorological changes was also investigated. In pursuing this, electrical measurements combined with meteorological observations were expected to offer additional insights into the physical processes causing the meteorological changes. Consequently quantitative studies of atmospheric electrification exist from the eighteenth century onwards: many are reviewed by Chauveau (1925).

### 1.2.1 Eighteenth century

Formal study of atmospheric electricity can usefully be regarded to have begun in 1750, with the pioneering thunderstorm experiments of Franklin in Philadelphia and D'Alibard in Paris (Fleming, 1939). Thunderstorm and lightning research still remain a central aspect of atmospheric electricity, but it is the awareness of electrification present in non-thunderstorm conditions, arising from the thunderstorm research, which is important in this paper. Lemonnier detected such electrification in clear air in 1752 (Chalmers, 1967) and Canton measured electrical changes arising from non-lightning producing clouds (Herbert, 1997). Other pioneering eighteenth century studies also retain some relevance: Beccaria (1775) measured daily variations in atmospheric electricity for a period of several years, including the existence of positive electrification on fine days, and noted the effect of fog on changing the electrical parameters. Provoked to further investigations by these findings, Read (1791) pursued a series of daily observations in London, using a vertical antenna connected to a pith ball electrometer. There are many examples of positive electrification during fair weather conditions. In his summary of two years of continuous observation from May 1789 to May 1791, Read recorded 423 days of positive electricity, 157 days of negative electricity and 106 days on which the electrification was sufficient to produce sparks (Read, 1792).

As well as Beccaria, de Saussure found diurnal variations in atmospheric electricity in European measurements made between 1785 and 1788. De Saussure records (Chauveau, 1925):

> *In winter, the season during which I have the best observations of serene[1] electricity...the electricity undergoes an ebb and flow like the tides, which increases and decreases twice in the span of twenty-four hours. The times of greatest intensity are a few hours after sunrise and sunset, and the weakest before sunrise and sunset.*

This serves to illustrate the readiness with which a diurnal variation could be observed with early instrumentation: although any physical interpretation now can only be speculative, the double diurnal cycle reported may have been related to variations in local smoke pollution.

## 1.2.2 Nineteenth century

Several sets of surface quantitative atmospheric electrical data exist from the nineteenth century, mostly of the vertical Potential Gradient (section 1.3.1). In the UK, systematic measurements of the Potential Gradient were made at Kew Observatory near London between 1844-7 by Sir Francis Ronalds (Ronalds, 1844; Scrase, 1934), using a lantern probe and straw electrometers[2]. A double diurnal cycle of the atmospheric electric field was found, which remained present at Kew beyond the end of the nineteenth century (Scrase, 1934). From 1861-1872, Wislizenus, a physician, made atmospheric electrical observations in St Louis, Missouri, which were subsequently analysed for solar cycle effects (Bauer, 1925). Lord Kelvin made major improvements in the instrumentation in the mid-nineteenth century. His automatic recording apparatus for electric field was installed at Kew in 1860 (Scrase, 1934). Measurements using the Kelvin apparatus from 1862-1864 at Kew are reported by Everett (1868), and the electrical measurements at Kew are available almost continuously from 1898 to 1981 (Harrison, 2003a). Many other European measurement stations operated in the nineteenth century (see also Table 2), and long series of observations exist from Brussels (1845-70), Perpignan (1886-1911) as well as Florence, Greenwich and Paris. The effect of smoke pollution on these urban measurements was often substantial and could dominate the diurnal variation; consequently the electrical observations offer an indirect method by which the diurnal smoke variations themselves can be recovered at high temporal (and in some cases hourly) resolution (Mössinger, 2004).

### 1.2.3 Twentieth century

Balloon ascents at the end of the nineteenth century and beginning of the twentieth century provided data on the vertical profiles of atmospheric electrical parameters, generally finding a decrease in the vertical electric field's magnitude with height (Chauveau, 1925; Chalmers, 1967). Balloon and airship measurements also showed variations in the electric field across the edges of stratiform clouds (Everling and Wigand, 1921; Wigand, 1928). Other developments in atmospheric electricity in the first part of the twentieth century arose from the established observations and campaigns, such as those in Paris (Chauveau, 1925) and at Kew. Following the discovery of radioactivity by Becquerel, V.F.Hess discovered cosmic rays, which provide an increase in ionisation with increasing height in the atmosphere (Hess, 1928). Gerdien (1905) designed a cylindrical electrode instrument to determine air conductivity, and the interactions of aerosols, clouds and ions became the formal study of fair weather atmospheric electricity. Systematic observations on the electrification of large ions (*i.e.* small radius aerosols, condensation nuclei and cloud condensation nuclei) were made at P.J. Nolan's laboratory in Eire (O'Connor, 2000), including aerosol electricity experiments on heavily polluted Dublin air. In his early work, the Nobel laureate C.T.R. Wilson worked on fundamental problems of fair weather atmospheric electricity at Cambridge (Galison, 1997), and established regular measurement of the air-Earth current at Kew in 1909, continuing for almost 70 years (Harrison and Ingram, 2004). Related research groups continued at the Cavendish Laboratory and under J. A. Chalmers at Durham (Hutchinson and Wormell, 1968) until the early 1970s.

The defining work of the early twentieth century, however, was the set of exploratory voyages of the wooden geomagnetic survey vessel *Carnegie*, operated by the Carnegie Institute of Washington. Between 1909 and its destruction by fire in 1929, many sequences of standardised atmospheric electricity measurements were made around the world's oceans. Oceanic air is particularly suitable for fair weather measurements as it is usually remote from continental sources of aerosol pollution, and the characteristic universal variation of the atmospheric electric field is still known as the *Carnegie* curve from the data obtained. Together with pioneering measurement of the electrical conductivity of the stratosphere in 1935 using the balloon *Explorer II* (Israel, 1970) and C.T.R.Wilson's (1920, 1929) global circuit theory, the *Carnegie* data strongly contributed to defining the twentieth century paradigm of fair weather atmospheric electricity.

### *1.3    Surface measurements*

### 1.3.1 The potential gradient

The vertical atmospheric electric field, or *Potential Gradient*[3] (PG), is a widely studied electrical property of the atmosphere. In fair weather and air unpolluted by aerosol particles, diurnal variations in PG result from changes in the total electrical output of global thunderstorms and shower clouds. Even early PG measurements, such as those obtained by Simpson (1906) in Lapland, Figure 2, show a variation suggestive of that seen in more modern work. A common global diurnal variation results from a diurnal variation in the ionospheric potential $V_I$ (Mülheisen, 1977), which modulates the vertical air-Earth conduction current $J_z$ (figure 1) and, in the absence of local effects, the surface PG (Paramanov, 1971). $V_I$ and $J_z$ are parameters less prone to effects from local pollution and are therefore more suitable for global geophysical studies, but of which far fewer measurements have been obtained than of the surface PG (Dolezalek, 1972; Harrison 2003a).

## 1.3.2  Small ions and air conductivity

Small ions are continually produced in the atmosphere by radiolysis of air molecules. There are three major sources of high-energy particles, all of which cause ion production in air: radon isotopes, cosmic rays and terrestrial gamma radiation. The partitioning between the sources varies vertically. Near the surface over land, ionisation from turbulent transport of radon and other radioactive isotopes is important, together with gamma radiation from isotopes below the surface. Ionisation from cosmic rays is always present, comprising about 20% of the ionisation over the continental surface. The cosmic ionisation fraction increases with increasing height in the atmosphere and dominates above the planetary boundary layer. Cosmic ray properties are reviewed by Bazilevskaya (2000).

After the PG, air conductivity is probably the second most frequently-measured surface quantity in atmospheric electricity. The slight electrical conductivity of atmospheric air results from the natural ionisation, generated by cosmic rays and background radioisotopes. For bipolar ion number concentrations $n_+$ and $n_-$ the total air conductivity $\sigma$ is given by

$$\sigma = \sigma_+ + \sigma_- = e\,(\mu_+ n_+ + \mu_- n_-) \qquad (1)$$

where $\mu_+$ and $\mu_-$ are the average ion mobilities, with $\sigma_+$ and $\sigma_-$ the associated polar conductivities. Air conductivity is strongly influenced by aerosol pollution: in aerosol-polluted air it is considerably reduced by ion-aerosol attachment (Dhanorkar and Kamra, 1997). In both aerosol-free and polluted air, the PG and air conductivity have been found to be related by Ohm's Law

$$J_z = \sigma E_z \qquad (2)$$

for a vertical air-Earth conduction current density $J_z$ and PG of magnitude $E_z$.

Balloon-carried instruments have found a substantial vertical variation in the air conductivity. Using data from the *Explorer II* ascent (Gish and Sherman, 1936), Gish (1944) introduced the concept of the columnar resistance, as the resistance of a unit area column of atmosphere, found by integration of the conductivity with height. The columnar resistance $R_c$ is typically ~120 P$\Omega$m$^2$. The bulk of the resistance of a unit column of atmosphere arises from ion attachment to aerosol present in the atmospheric boundary layer; the conductivity steadily increases with increasing cosmic ray ionisation, up to the highly-conductive lower layers of the ionosphere.

### 1.3.3 Instrumentation for surface measurements

Surface instrumentation provides the majority of atmospheric electrical data available, and this Section summarises some of the classical techniques for measurement of the Potential Gradient, air conductivity and the vertical air-Earth current $J_z$.

Early PG measurements mostly used a potential probe (also known as a *collector*), schematically represented in figure 3 (a). The PG is derived from the measured potential $V$ at the height $z$ of the collector, and is typically 150V.m$^{-1}$ at 1m under fair weather conditions, although this varies between sites and between land and sea. Many different collectors have been employed in the past, including a burning fuse (Scrase, 1934), the water dropper originated by Lord Kelvin (Chalmers, 1967), a radioactive source (Israel, 1970) and a long horizontal wire antenna (Crozier, 1963; Harrison, 1997a). In all these approaches, the effective resistance between the air and the probe is reduced, either by increasing the area of the collector (long wire), or by introducing additional ions into the region of air adjacent to the collector (burning fuse, water dropper and radioactive source.)

Gerdien (1905) developed a method of measuring air conductivity using a cylindrical condenser. A voltage is applied between two cylindrical electrodes and the inner one is earthed via a sensitive ammeter, figure 3 (b). In the tube, ions of the same sign as the polarising voltage are repelled by the outer electrode, and move in the electric field to meet the inner electrode, where they cause a small current $i$. As long as the output current is proportional to the polarising voltage, indicating that a fixed fraction of air ions are collected by the central electrode, the polar conductivity $\sigma_+$ or $\sigma_-$ is given by

$$\sigma_\pm = \frac{i\varepsilon_0}{CV_\pm} \qquad (3),$$

where $C$ is the capacitance of the Gerdien device, and $V$ the polarising voltage established across the electrodes, of appropriate sign for the ions selected, and $\varepsilon_0$ is the permittivity of free space.

Measurement of the air-Earth current and potential gradient was combined in an instrument developed by C.T.R.Wilson, figure 3 (c). Wilson's apparatus (Wilson, 1906) consisted of a portable gold leaf electrometer, connected to a horizontal, circular collecting plate, surrounded by an earthed guard ring. The apparatus also included a battery and a variable capacitor (or *compensator*), by which the potential of the collecting plate could be changed, and a brass cover for the collecting plate. From a sequence of covering and uncovering the plate, and timing the change in potential, the air-Earth current density $J_z$ and mean PG could be found. (A further important aspect of the apparatus was the compensation of displacement currents, which are produced as a result of short-term PG fluctuations.)

In modern applications, PG measurement is usually made using mechanical field machines (Chalmers, 1967), more commonly called *field mills*. Field machines generally offer a more rapid time response and dynamic range (Chubb, 1990; MacGorman and Rust, 1998) but may require maintenance with atmospheric exposure. The Gerdien instrument continues to provide the physical basis on which modern ion counting instruments are still designed (Blakeslee and Krider, 1992; Hõrrak *et al*., 1998), although computer control of the bias voltage and current measurements is becoming more common (Aplin and Harrison, 2000, 2001; Holden, 2004). Direct measurement of the air-Earth current has been achieved using a split-hemisphere collector, suspended above the surface, within which the current amplifiers are housed (Burke and Few, 1978). This has also proved durable under Antarctic conditions (Byrne *et al.*, 1993). A further effective method of air-earth current measurement is the long wire antenna collector (Tammet *et al.*, 1996).

One characteristic of fair-weather atmospheric measurements is that the signal voltages are present at high source impedance, and therefore sensitive electrometry techniques are required. The necessary electrometer dynamic range has usually required a high value potential divider, but the resistor technology introduces thermal stability issues. This problem can be circumvented by circuit design (Harrison, 1996, 2002a), at sufficiently low cost that the equipment is disposable and therefore suitable for use with non-recovered balloon-carried instrumentation (Harrison, 2001). In the related area of ultra-low electric current measurements, compensation of temperature-dependent errors (Harrison and Aplin, 2000a) and power line interference (Harrison, 1997b) can now be achieved with stable current references (Harrison and Aplin, 2000b). High integrity of insulation is always necessary, and, although the data logging can readily be automated, removal of cobwebs and general contamination requires regular maintenance.

## 2   The global circuit

The classical concept of a global electrical circuit was suggested by C.T.R.Wilson (Wilson, 1920, 1929), following the first reports of a universal time diurnal variation in the electrical parameters (Hoffman, 1924), and knowledge of the conductive upper regions of the atmosphere inferred from radio wave investigations (Fleming, 1939). The global atmospheric electrical circuit arises from a balance between charge generation in disturbed weather regions and a globally distributed current of small ions flowing in fair weather regions (Rycroft *et al.*, 2000). This is shown schematically on the left-hand side of figure 1. This simplified representation is generally understood as a low latitude direct current (DC) atmospheric electric circuit as, at high latitudes, polar cap potentials contribute additional currents.

As well as providing a global equipotential region, the ionosphere provides the upper boundary of a spherical resonant cavity. Excitation of the earth-ionosphere cavity by lightning-generated electromagnetic energy causes alternating current (AC) effects, notably the Extremely Low Frequency (ELF) Schumann resonance at ~8Hz (*e.g.* Rycroft *et al.*, 2000), the intensity of which is closely related to the total amount of planetary lightning. Because the DC circuit responds to current generation from *both* shower clouds and thunderstorms, differences arise between the DC parameter of the ionospheric potential and the lightning-driven Schumann resonance amplitude. Existence of an ELF resonance on other bodies of the solar system seems likely: a fundamental ELF frequency of 11-15Hz has been postulated for Titan (Morente *et al.*, 2003), and 13-14Hz for Mars (Sukhorukov, 1991). Since planetary (or satellite) ELF resonances require the combination of an upper atmospheric conducting layer and sources of atmospheric electrical discharges, detection of ELF indicates the existence of an AC atmospheric electrical circuit. For a DC circuit to be present in the terrestrial sense, additional evidence of current flow is required.

Table 1 provides a summary of properties of the terrestrial atmospheric electric circuit parameters.

**Table 1. Typical parameters of the fair weather atmospheric electric circuit**

| | | |
|---|---|---|
| surface Potential Gradient (PG) | $\mathbf{E}$ | 120 V.m$^{-1}$ |
| vertical conduction current density | $\mathbf{J_z}$ | 3 pA.m$^{-2}$ |
| air conductivity | $\sigma$ | 20 fS.m$^{-1}$ |
| | | (mean at sea level) |
| ionospheric potential | $V_I$ | 250 kV |
| total resistance of the atmosphere | $R_T$ | 230$\Omega$ |
| total air-Earth current | | 1800A |
| columnar resistance | $R_c$ | 120 P$\Omega$.m$^2$ |

### 2.1 *Carnegie diurnal variation*

Section 1.3.1 introduced the diurnal variation in PG identified in data obtained during the voyages of the *Carnegie*. As well as the surface PG, the universal diurnal or *Carnegie curve* appears in the other global circuit parameters, notably the ionospheric potential (Section 2.2.1) (Mülheisen, 1977) and air-Earth current (Section 2.2.2) (*e.g.* Adlerman and Williams, 1996). A positive correlation between the Carnegie curve and diurnal variation in global thunderstorm area was discovered by Whipple, by summing the diurnal variations in thunderstorm area for each of Africa, Australia and America (Whipple, 1929; Whipple and Scrase, 1936). The thunderstorm areas were estimated using thunderday[4] statistics from meteorological stations, compiled by Brooks (1925). Figure 4 (a) shows the diurnal variations in surface PG and thunderstorm area: the correlation between the 24 data points is +0.93. The Carnegie curve is tabulated in Israel (1973), and its values are also given in Table 2, together with the thunderstorm areas of Whipple and Scrase (1936) from thunderday data.

**Table 2. Hourly values of the Carnegie curve Potential Gradient and global thunderstorm area found by Whipple and Scrase (1936) (from Israel, 1973)**

| hour GMT (or UT) | PG(V/m) | % | Thunderstorm area ($10^4$km$^2$) | % |
|---|---|---|---|---|
| 1 | 112 | 87 | 161 | 85 |
| 2 | 110 | 86 | 154 | 81 |
| 3 | 109 | 85 | 152 | 80 |
| 4 | 111 | 86 | 155 | 81 |
| 5 | 113 | 88 | 161 | 85 |
| 6 | 116 | 90 | 169 | 89 |
| 7 | 119 | 92 | 175 | 93 |
| 8 | 120 | 93 | 179 | 95 |
| 9 | 121 | 94 | 181 | 96 |
| 10 | 122 | 95 | 197 | 99 |
| 11 | 124 | 96 | 196 | 103 |
| 12 | 129 | 100 | 205 | 108 |
| 13 | 135 | 105 | 215 | 113 |
| 14 | 139 | 108 | 220 | 116 |
| 15 | 142 | 110 | 216 | 114 |
| 16 | 144 | 112 | 212 | 112 |
| 17 | 147 | 114 | 214 | 113 |
| 18 | 151 | 117 | 217 | 114 |
| 19 | 154 | 120 | 218 | 115 |
| 20 | 151 | 117 | 215 | 114 |
| 21 | 143 | 111 | 207 | 114 |
| 22 | 135 | 105 | 197 | 104 |
| 23 | 128 | 99 | 182 | 96 |
| 24 | 122 | 95 | 171 | 90 |
| *mean* | 129 | | 189 | |

In a more modern analysis of available global data, Paramanov (1971) confirmed a Universal Time diurnal variation in both the PG and air-Earth current, for all seasons. Later analysis of the global thunderstorm area variation, however, has produced less conclusive results than Whipple and Scrase (1936). Krumm(1962) analysed World Meteorological Organisation records by latitude band and season and found a global maximum around 14UT; Dolezalek (1972) found a diurnal maximum at 16UT. This paradox between the thunderstorm area and universal diurnal variation data can be resolved if, as well as substantial African thunderstorm contributions to the global circuit around 14-15UT, there are large current contributions from South American shower clouds around 20UT (Williams and Sátori, 2004).

Figure 4(b) presents the original Universal Time variation data in an alternative way. This figure has been drawn to emphasise the effect of differences in the amplitude of the diurnal curves of thunderstorm area and PG in Figure 4 (a). It is clear that, as the thunderstorm area tends to zero, there is a non-zero intercept. This implied existence of the PG in the absence of continental thunderstorms was originally thought to be due to the presence of oceanic thunderstorms (Whipple, 1929), neglected in the thunderstorm area evaluation, for which an arbitrary offset, with no diurnal variations, was added (Williams and Heckman, 1993). In a review of earlier work by Wormell (1953), the negative charge of the Earth's surface was estimated to arise from the balance between fair weather current (60C.yr$^{-1}$.km$^{-2}$) and precipitation (30 C.yr$^{-1}$.km$^{-2}$), lightning (-20 C.yr$^{-1}$.km$^{-2}$) and point-discharge currents (-100C.yr$^{-1}$.km$^{-2}$), indicating that lightning was not the dominant charging agent. Williams and Heckman (1993) argued that the corona and precipitation current contributions to the conduction currents at the surface are the principal source of charging of the global circuit, emphasising the role of electrified shower clouds as originally suggested by Wilson (1920). In support of this view, recent satellite lightning observations have shown that there is little marine lightning (Christian, 2003). Furthermore, variations in global lightning alone are not quantitatively sufficient to account for the variations in the global circuit. In a study comparing ELF measurements (used as a proxy for global lightning) with the PG measured in the Antarctic, the mean variations were similar but the hourly departures were less well correlated (Füllekrug *et al.*, 1999). Füllekrug *et al.* (1999) reported that the contribution of lightning to the global circuit to was ~40 ± 10%.

Surface PG observations show that the Carnegie curve has been largely unchanged during the twentieth century. At Eskdalemuir (Scotland) and Lerwick (Shetland), a long series of hourly PG measurements was made in clean air, under standardised conditions (Harrison, 2003a). The records began in Scotland in 1908 and at Shetland in 1926, continuing until both stations ceased atmospheric electrical measurements in the early 1980s. Figure 5(a) shows averaged daily variations of PG obtained at Lerwick 1968-1973, compared with the *Carnegie* curve and the contemporary ionospheric potential changes found by Mülheisen (Budyko, 1971). There are qualitative similarities with the PG variation found by Simpson at Karasjok in 1903, Figure 2, again indicating some temporal consistency in the diurnal variation across the twentieth century.

Averaging many fair weather days of surface PG data is usually necessary for the Carnegie curve to appear, both because of local aerosol influences and also because day-to-day variability in the global circuit–modulating thunderstorms may prevent the standard variation occurring every day. This is supported by the original measurements (Torreson, 1946) on Cruise VII of the *Carnegie* (1928-1929), in which diurnal variations close to, but different from, that subsequently identified as the standard variation were occasionally observed on fair weather days, without any averaging. On at least one occasion during Cruise VII, a similar, but non-standard diurnal variation was observed simultaneously on the *Carnegie* and at Eskdalemuir (Harrison, 2004a). Figure 5(b) shows a series of six-hourly PG values obtained on the *Carnegie* in September 1928, with measurements made on undisturbed days at Eskdalemuir. From the 12 to 17 September 1928, there was a significant correlation ($r = 0.56$ for 20 values) between the PG observed on the *Carnegie* and at Eskdalemuir, with particularly close agreement on the 15 and 16 September. The absolute values of PG between the two sets of measurement differ, however, which is likely to be due to local factors, such as the presence of aerosol or local geological properties influencing the surface ion production and the associated air conductivity.

Figure 6 shows average diurnal variations in potential in pioneering measurements made by B. Chauveau in Paris (Chauveau 1893a, 1893b, 1925), in the summer of 1893. These were made both at the surface, and at the top of the Eiffel Tower, using the Kelvin water dropper apparatus with autographic recording (Harrison and Aplin, 2003). There are qualitative differences between the two sets of measurements. In particular, the surface measurements have two maxima, in common with those also found in London in the nineteenth century (Scrase, 1934), which were due to the similar diurnal variations in smoke pollution (Harrison and Aplin, 2002). The contrast between the lower and upper measurements is mentioned by Chauveau (1925):

> *In observations made during the summer and autumn months, we noted that, rather than the strongly marked double period oscillation found at the lower station, the variation simultaneously obtained at the upper station has a very different character when averaged, clearly showing a single oscillation. After the morning minimum, occurring exactly at the same time at the two stations, the potential seen around the top of the Eiffel Tower increases steadily until an evening maximum which occurs an hour earlier than the maximum observed at the base station.*

The single maximum in the data from the top of the Eiffel Tower, and the increased amplitude over the Carnegie curve, result from a combination of the global circuit modulation nocturnally with the diurnal variation in transport of polluted urban boundary layer air past the sensor (Harrison and Aplin, 2003).

## 2.2    DC aspects of global circuit

In the DC global circuit, the ionospheric potential $V_I$ causes the vertical conduction current $J_z$ to flow through the combined vertical resistance of the troposphere and stratosphere (see also Section 1.3.2.). From the measured air-Earth current and planetary surface area, an estimate of the total current flowing in the global circuit is 2000A. Assuming a simple circuit and applying Ohm's Law, the global electrical resistance $R_T$ is about 230Ω. The lower part of the troposphere, the boundary layer, contributes a significant part of the vertical resistance. $R_T$ can be found by integrating the variation of air conductivity with height as

$$R_T = \frac{1}{4\pi R^2} \int_0^\infty \frac{dz}{\sigma(z)} \qquad (4)$$

where $R$ is the Earth's radius. Variations in $R_T$ arise from changes in cosmic ray ionisation, clouds and aerosols: it may therefore reflect long-term climate changes or changes in the abundance of volcanic aerosol. The concentric sphere system formed by the conducting ionosphere surrounding the planet has a finite capacitance, with a *RC* time constant of ~10 minutes. The continued existence of an atmospheric electric field, despite this rapid discharging timescale, was the original indication that electrical generation processes are continuously active (Chalmers, 1967).

### 2.2.1 The ionospheric potential

The ionospheric potential $V_I$ is the potential of the lowest conducting regions of the ionosphere, reckoned with respect to a zero potential at the surface. It is obtained from vertical soundings of the variation of PG with height, $E(z)$ using a sensor carried on an ascending balloon or aircraft, and the total potential calculated by integration,

$$V_I = \int_0^{z_I} E(z)dz \qquad (5).$$

Because the air conductivity increases substantially in the upper troposphere, the value of $V_I$ does not have a high sensitivity to the upper limit, $z_I$, of the ascent (Markson, 1976). Many intermittent measurements of $V_I$ exist from the late 1950s (Imyanitov and Chubarina, 1967; Markson, 1976, 1985; Markson *et al.*, 1999). Those of Mülheisen (Budyko, 1971) obtained in Germany between 1959 and 1972 are notable for providing important support for the global circuit concept from the common simultaneous variations in $V_I$ found at different locations (Mülheisen, 1971). Figure 7 shows the comparison between simultaneous $V_I$ soundings made in Weissenau, Germany, and over the Atlantic from the research ship *Meteor*. In the semi-continuous measurement period between 17 March and 2 April 1969, there was both quantitative and qualitative agreement in the $V_I$ values obtained (correlation coefficient $r = 0.82$, $N = 15$ values). Figure 7 also includes the measurements of surface PG obtained at Lerwick at the same UT hour as the soundings of Mülheisen, for fair weather or non-precipitating periods. There is also clearly good agreement between the determination of $V_I$ (averaged from the two simultaneous soundings) and the Lerwick PG data, ($r = 0.87$, $N= 10$).

Increases in surface temperature leading to more vigorous convection have been quantitatively linked with increases in $V_I$ (Price, 1993), offering a possible indirect method of monitoring surface temperatures. For example, a positive correlation between three different global temperature datasets and $V_I$ (Markson and Price, 1999), supporting a relationship between warmer temperatures and an increased $V_I$.

### 2.2.2 The conduction current

The conduction current was defined in Section 1.3, and is related to the resistance $R_c$ of a unit area column of atmosphere and ionospheric potential by

$$J_z = \frac{V_I}{R_c} \qquad (6)$$

where $J_z$ is the conduction current density. At the surface, $J_z$ can be determined as the current flowing from the atmosphere to ground, the *air-Earth current density*, of which a long series (1909-1979) of approximately weekly measurements exists from Kew near London using the Wilson apparatus (Harrison and Ingram, 2004). Continuous measurements of the air-Earth current in clean air (at Hawaii) were made with other atmospheric electrical parameters during 1962 (Cobb, 1968).

Measurement of $J_z$ will not provide a method of monitoring changes in the global circuit if the columnar resistance varies locally (Markson, 1988). However, if $R_c$ is constant, $J_z$ may be closely linked to $V_I$. Evidence for this comes from averaging measurements of $J_z$ which, in oceanic air, also show the Carnegie curve (Gringel *et al.*, 1986). In balloon soundings, $J_z$ has been found to be constant with height above the boundary layer (Gringel, 1978).

### 2.3  AC aspects of the global circuit

The production of ELF radiation by lightning offers a direct method of monitoring the temporal variation of terrestrial lightning. Williams (1994) has shown that lightning flash rates are proportional to temperature, and it has been suggested that the Schumann (ELF) resonance could offer a method of monitoring global temperature (Williams, 1992). A significant advantage of using the Schumann resonance over the ionospheric potential (Section 2.2.1) to monitor global thunderstorms is that ELF measurements can be made at the surface in electromagnetically quiet regions and, therefore, that no balloon soundings are required. As a result, short-term (seconds to hourly) temporal changes can be monitored (Märcz *et al.*, 1997). Study of the AC (*i.e.* lightning-related) aspects of the global circuit is also unaffected by continental aerosol, which contributes variability in the columnar resistance of the DC global circuit. Temperature effects on the global circuit are considered further in Section 3.2.4.

#### 2.3.1  Satellite observations of lightning

Observations of lightning can be made by meteorological lightning detection ("sferics") networks, or obtained from satellite, which view a large proportion of the thunderstorm-producing regions of the planet. Early satellite observations of the distribution of midnight lightning (Orville and Henderson, 1986) showed that there was more lightning over land and that the lightning frequency had a maximum in the northern hemisphere summer. More recent sensors, including the NASA Optical Transient Detector (OTD) and Lightning Imaging Sensor (LIS), are able to detect lightning under daylight conditions (Christian, 1999; Christian *et al.*, 2003). Detailed climatologies of the planetary lightning distribution now exist. These show that 78% of all lightning occurs between 30°S and 30°N, with Africa the greatest source region of lightning and that there is a global mean land to ocean lightning flash ratio of 10:1 (Christian, 2003).

### 2.4  Non-diurnal modulations

Modulation of the global atmospheric electrical circuit occurs on both diurnal and seasonal cycles. The diurnal cycle is sufficiently well established that it can be used for preliminary identification of data of possible global validity (Märcz and Harrison, 2003; Harrison, 2004b). There are fewer observations with which to identify seasonal, intra-seasonal (Anyamba *et al.*, 2000) and longer-term changes.

#### 2.4.1  Seasonal variation

Satellite observations of visible lightning provide high-resolution data on the seasonal variation of lightning. The northern hemisphere lightning, associated with the majority of the planetary land area, contributes the majority of the global lightning. The global flash rate has a maximum during the northern hemisphere (NH) summer (Figure 8a). Early measurements of surface Potential Gradient, however, show a maximum in the NH winter. The 1903/4 Lapland PG data (Figure 2), for example, show what was originally assumed to the seasonal variation; many other early NH datasets show this (Chauveau, 1925).

By comparing northern and southern hemisphere PG data, Adlerman and Williams (1996) argued that variations in NH continental smoke pollution were the origin of the NH winter PG maximum. The pollution shows a maximum in the winter; consequently the PG measurements may not be representative of the global circuit's seasonality. Reanalysing all the data obtained on the *Carnegie* voyages, and those of the *Maud*, another early twentieth century survey vessel, Adlerman and Williams (1996) found a maximum in PG in June/July, consistent with the maximum in lightning occurrence.

In an analysis of data obtained at the Marsta station in Sweden between 1993-1998, Israelsson and Tammet (2001) show the strongest correlation between the local PG variation and the Carnegie curve in the NH winter ($r = 0.96$). A similar behaviour has been found in data from the UK and Hungary (Märcz and Harrison, 2003), and in data from the Bavarian Alps (Harrison, 2004b). Since these studies consider the shape of the Carnegie curve, rather than the absolute magnitude, these observations could suggest that either the DC global circuit is at its strongest during the winter, or short-term changes are particularly small, or a combination of the two factors. Short-term (hourly) fluctuations may arise in the summer measurements from the effects of local thunderstorms. Figure 8 (b) shows the PG data from cruise VII of the Carnegie, plotted as monthly averages. There are variations in the shape of the curve from summer to winter, and in the timing of the maxima and minima. Important though these data are in the development of atmospheric electricity, they span less than two years, and may not be representative of the average seasonal variation.

Although the seasonality in lightning is well-established and drives the AC global circuit, Section 2.2 emphasised that lightning is not the sole source of variability in the DC global circuit. The seasonality in the AC and DC circuits may not therefore be exactly linked. Furthermore, changes in the columnar resistance, caused by the seasonal variations in cloud and aerosol, will affect currents flowing in the DC global circuit. These factors will need to be considered further in definitively determining the seasonal variation, in the absence of direct long-term monitoring of the DC global circuit.

### 2.4.2 Cosmic ray modulation

A small observed modulation in $V_I$ has been linked to cosmic ray variations arising from the solar cycle (Markson, 1981): cosmic ray ionisation modifies the atmosphere's columnar resistance. $V_I$ increased when an increase in cosmic rays occurred. Markson (1981) showed that, for the increase in cosmic rays to *increase* $V_I$, the change has to influence the *charging* part of the global electric circuit.

The cosmic ray effect on $V_I$ was originally identified from the solar modulation of low energy cosmic rays, which occurs from the solar variations on the eleven-year cycle. Interaction between the solar wind and cosmic rays leads to an inverse phase relationship between solar activity and cosmic rays arriving at Earth. A more active Sun therefore reduces the cosmic ray flux penetrating down into the Earth's lower atmosphere, and the steady increase in solar activity during the twentieth century has lead to a secular decline in cosmic rays (Carslaw *et al.*, 2002). The expected global circuit response may have been identified in surface measurements of PG (Harrison, 2002b; Märcz and Harrison, 2003), although aerosol changes have also been suggested to cause the effect (Williams, 2003; Harrison, 2003b). A twentieth century decline is, however, also apparent in oceanic measurements of PG (Harrison, 2004a), which supports a global circuit change (see also Section 3.1.1.) A decrease in the air-Earth current at Kew (Harrison and Ingram, 2004) and in the PG in mountain air (Harrison, 2004b) is apparent in the 1970s.

# 3 Changes in the global circuit and climate

## 3.1 *Reconstruction of past global circuit variations*

The pioneering measurements described in Section 1.2 are of historical importance in understanding the development of atmospheric electricity, but they have recently found a new use in extracting information on past urban air pollution conditions. Another atmospheric use for the early measurements is emerging, as studies of past climates and more recent atmospheric changes are increasingly used to provide a comparative basis for research on climate change. In the case of global air temperature for example, early observations have been used to quantify past temperature changes, globally or regionally. The longest such temperature series, the Central England Temperature series, extends back to 1698 (Manley, 1953). Such an approach of instrumental reconstruction should, in principle, be possible for the global atmospheric electrical circuit. Because of its global nature in fair weather conditions and clean air, rather fewer instrumental measurements would seem likely to be needed than for computing the global temperature. In many cases the experimental work is recorded in sufficient detail for the instruments to be simulated and estimates of their reliability made. Even so, major difficulties remain in attempting to reconstruct the past state of the global electrical circuit, including the absolute calibration of early instrumentation, changes in the exposure of the measurement sites and gaps for which no data are available.

A comparable approach in paleoclimate studies is the use of proxy methods to estimate changes in past conditions. A suitable proxy has a known relationship to the physical quantity actually required, and its variations are known for the period of interest, usually when the physical quantity is not known directly. Proxy methods have been used to infer long-term variations in solar output, from records of cosmogenic isotopes, which respond inversely to solar changes. Central to both the instrumental and proxy methods for a global atmospheric circuit reconstruction is, however, the availability of long-term, globally-valid measurements.

### 3.1.1 Early measurements

Table 3 lists the locations of early surface measurement stations and Figure 9 illustrates the periods of operation of a subset, including the UK sites. As a first stage in selecting data of possible global significance, the diurnal variations described in Section 1.1.1 may be used to reject data primarily influenced by local pollution. The Carnegie curve provides one method for identifying periods of data in which global circuit signals may be present (Harrison, 2002b) or, from the absence of the universal diurnal cycle, when the PG is dominated by local influences (Harrison and Aplin, 2002). This approach is necessary when little or no additional information beyond the PG measurements is available. The existence of a similar variation to the Carnegie curve is not, however, a sufficient condition for global circuit monitoring, as it may have arisen from local changes having the same phase variation by chance. Identification of global circuit variations requires additional corroboration with simultaneous atmospheric electrical changes found at other sites (Harrison, 2004a, 2004b).

A long-term decline in the average annual PG was first observed in data from Eskdalemuir in rural Scotland (Harrison, 2002b). In urban air at Kew, the data are strongly affected by local smoke pollution from London, but there is an annual cycle in air pollution, with the smoke effects at Kew smallest in the summer (Harrison and Aplin, 2002). When the annual PG measurements from Eskdalemuir are compared with the Kew June PG measurements in the first half of the twentieth century, a similar rate of decline emerges, see figure 10a. (This is also found in the PG data from Lerwick, which used identical instrumentation, for the summer months of June, July and August.) Williams (2003) has suggested that a steady decrease in continental aerosol could explain the decrease in the PG found at Eskdalemuir. Although this is physically plausible, Harrison (2003b) argued that, before the UK Clean Air Acts of the 1950s, systematic *decreases* in smoke were unlikely, as the Clean Air legislation resulted from more frequent and more damaging air pollution episodes. Furthermore, there are air-Earth current measurements for a similar period in the UK, using the Wilson apparatus (section 1.3.3). This apparatus also measured the air conductivity, which is the physical parameter modulated by smoke pollution. (For constant $J_z$, the PG and conductivity are proportional, from equation 2.) A greater decline in the air-Earth current than in the positive air conductivity is found at Kew in June, from the available data from 1932-1948 (Harrison and Ingram, 2004). This supports the decline in the measured PG being due to air-Earth current (*i.e.* global circuit changes), rather than air conductivity (*i.e.* local pollution) changes.

**Table 3. Early quantitative surface measurements of the Potential Gradient**

| Station | start | finish | annual data | | daily data | variation type | Source |
|---|---|---|---|---|---|---|---|
| | | | mean PG V/m | %variation | %variation | | |
| Melbourne | 1858 | 1860 | 145 | 52 | 86 | | 1 |
| Sodankyla | 1882 | 1883 | | 85 | 29 | S | 3 |
| Florence | 1883 | 1886 | | 55 | 31 | S | 3 |
| Lisbon | 1884 | 1886 | | 30 | | | 3 |
| Perpignan | 1886 | 1888 | | 48 | | | 3 |
| Batavia (at 2m) | 1887 | 1890 | | 93 | | | 3 |
| Batavia (at 7m) | 1890 | 1895 | | 47 | | | 3 |
| Greenwich | 1893 | 1896 | | 83 | 44 | D | 3 |
| Batavia | 1890 | 1900 | | 44 | 121 | | 4 |
| Paris Central | 1893 | 1898 | 175 | 40 | 57 | D | 1 |
| Paris Eiffel (at 1.7m spacing) | 1893 | 1898 | 4010 | slight | 45 | S | 1 |
| Helsinki | 1895 | 1897 | | | 100 | S | 1 |
| Tokyo | 1897 | 1901 | | 91 | | D | 3 |
| McMurdo Sound | 1902 | 1903 | 93 | 68 | | | 4 |
| Karasjok | 1903 | 1904 | 139 | 86 | 79 | S | 2 |
| Trieste | 1902 | 1905 | 73 | 34 | 118 | | 1 |
| Munich | 1905 | 1910 | 168 | 96 | 79 | | 4 |
| Petermann Island | | | 164 | 116 | 56 | | 4 |
| Davos | 1908 | 1910 | 64 | 106 | 69 | | 4 |
| Kremsmünster | 1902 | 1916 | 105 | 75 | 57 | | 4 |
| Vassijaure | 1909 | 1910 | 89 | 124 | 56 | S | 1 |
| Cape Evans | 1911 | 1912 | 87 | 46 | 35 | S | 1 |
| Helwan | 1909 | 1914 | 150 | 36 | 41 | | 4 |
| Upsala | 1912 | 1914 | 70 | 84 | 71 | D | 4 |
| Ebeltofthann(Svalbard) | 1913 | 1914 | 95 | 80 | 17 | | 1 |
| Potsdam | 1904 | 1923 | 202 | 63 | 36 | | 4 |
| Edinburgh | | | 167 | | 79 | | 5 |
| Eskdalemuir | 1914 | 1920 | 263 | 61 | 42 | S | 4 |
| Tortosa | 1910 | 1924 | 106 | 35 | 54 | | 4 |
| Carnegie | 1915 | 1921 | 124 | 16 | 35 | | 4 |
| Aas | 1916 | 1923 | 104 | 101 | 44 | | 4 |
| Val Joyeux | 1923 | 1924 | 90 | 70 | 53 | | 1 |
| Wahnsdorf | 1924 | 1926 | 178 | 79 | 57 | | 1 |
| Frankfurt (Feldbergstrasse) | 1928 | 1931 | 146 | | 46 | | 1 |
| Tromso | 1932 | 1933 | 104 | 28 | 50 | | 1 |
| Scoresby Sound | 1932 | 1933 | 71 | 50 | 45 | | 1 |
| Fairbanks | 1932 | 1933 | 97 | 30 | 38 | | 1 |
| Chambon-la-Fôret | 1942 | 1944 | 92 | 80 | 50 | | |
| Heidelberg | 1957 | 1958 | 129 | 112 | 28 | | 1 |

*Sources*: [1] Israel (1970); [2] Simpson (1906); [3] *Atmospheric Electricity* entry in *Encyclopedia Britannica* (1911); [4] Benndorf (1929); [5], Chree (1915). *S* and *D* denote single and double oscillation diurnal variations, respectively, where known.

### 3.1.2  Proxies for changes in the global circuit

Other than early direct measurements, one of the longest sources of local atmospheric electrical data is the regular meteorological reporting of thunderdays, which formed the basis of early (Brooks, 1925) and later (Krumm, 1962; Dolezalek, 1972) studies on the global circuit. Thunderdays have been investigated for possible solar influences (Brooks, 1934): there are close statistical correlations apparent in some regions (Stringfellow, 1974; Schlegel *et al.*, 2001). The widespread recording of the occurrence of thunder has suggested a possible route for investigating global electrical changes (Changnon, 1985), but the trends in different continental areas vary, suggesting that local factors may dominate. On millennial timescales, historical evidence from China indicates three major peaks in winter thunderstorm activity in 200-600AD, 1100-1300AD and 1600-1800AD (Wang, 1980).

From the perspective of understanding global circuit changes, a more quantitative proxy is desirable. Since the global circuit requires cosmic ray ionisation to cause ion production and with it the conductivity of air, cosmic ray measurements provide information on an important physical aspect. Direct measurements of cosmic ray secondary particles with surface neutron counters began in the 1950s, with ionisation chambers collecting the data before then, such as those carried on the *Carnegie* cruises. Before the first decade of the twentieth century, proxies for cosmic rays provide indirect measurements of the atmospheric ion production rate. Such cosmic ray proxies indicate possible changes in the columnar resistance of air within the global circuit (Section 2.4.2), and may permit the calibration of modern $V_I$ and PG data to obtain past estimates of their variations. Estimates of cosmic ray variation during the Phanerozoic (the past 500 million years) have been claimed from meteorite data (Shaviv and Veizer, 2003). Over shorter timescales (the past 100,000 years), one proxy for global cosmic ray intensity is the cosmogenic nuclide beryllium-10, which is generated by spallation reactions of cosmic rays in the stratosphere. A long archive of Be-10 data exists from an analysis of deposits in the Greenland ice core, which readily shows the 11 year solar cycle modulation of cosmic rays (Beer, 2000).

Evidence discussed in Section 3.1.1 suggests a global circuit change may have occurred during at least the first half of the twentieth century. Should this be solely due to changes in the columnar resistance from cosmic ray ionisation at the same time, it is possible that the Be-10 measurements could offer a proxy for variations in the atmospheric electrical circuit (Harrison, 2002b), beyond the period of the direct observations. A further reason for the possible usefulness of Be-10 is that it is formed in the upper part of the atmosphere, where cosmic ray ion production modulates the charging part of the global circuit. Figure 10 (b) shows the smoothed Be-10 variation from 1600 (Beer, 2002), calculated as a fraction of the mean ice core Be-10 concentration from 1900-1950. The relative PG variation shown in Figure 10 (a) is also plotted for the period 1900-1950 in Figure 10(b). Quantitative comparison requires the Be-10 relationship with atmospheric ion production to be calculated, but there is a physical basis for expecting the global circuit and cosmic ray ion production to be linked through columnar resistance changes (Markson, 1981).

### 3.2 Consequence of changes in atmospheric electrification

#### 3.2.1 Electrical effects on cloud microphysics

There is a considerable variety in the sizes and abundance of aerosol particles, ice crystals and cloud droplets present in the atmosphere. The typical molecular cluster comprising an atmospheric small ion will have a diameter of less than one nanometre. Aerosols have diameters from 3nm to 10µm, and cloud and raindrops have diameters from 10µm to 1mm. The complexity of cloud microphysical processes results from both the different particle sizes and compositions in atmospheric clouds, and the phase changes between solid and liquid (Mason, 1971; Pruppacher and Klett, 1997).

Aerosol electrification in the atmosphere occurs from ion-aerosol attachment, facilitated by ion transport in electric fields (Pauthenier, 1956), by diffusion (Gunn 1955; Fuchs, 1963; Bricard, 1965; Boisdron and Brock, 1970). Chemical asymmetries in the ion properties generally result in a charge distribution on the aerosol, which, although the distribution may have a small mean charge, does not preclude the existence of transiently highly charged particles within the ensemble. In the special case of radioactive aerosols, the charge arises from a competition between the self-generation of charge within the particle, and ions diffusing to the particle (Clement and Harrison, 1992; Gensdarmes *et al.*, 2001). The aerosol charge distribution can affect aerosol coagulation rates (Clement *et al.*, 1995), which in turn may modify the particle size distribution.

Recent review papers discuss the detailed electrical microphysics which links cloud processes with atmospheric electrification (Harrison, 2000; Carslaw *et al.,* 2002). Two aerosol-related topics identified are :

(1) Cosmic ray generated atmospheric ions have been shown theoretically to lead to the formation of atmospheric Condensation Nuclei (Yu and Turco, 2001), for which there is some experimental support (Wilkening, 1985; Harrison and Aplin, 2001).

(2) Aerosol electrification can modify the rate at which the aerosol particles are collected by water drops, which may affect the freezing of supercooled water drops. (Tinsley *et al.*, 2000; Tripathi and Harrison, 2002).

More quantitative discussions of these mechanisms, based on the theory already available, are given in Harrison and Carslaw (2003).

In the first case, atmospheric effects would occur through changes in ion production, leading in turn to changes in aerosol nucleation, ion-aerosol attachment rates, aerosol charge equilibration times and aerosol coagulation rates, ultimately modifying clouds. Direct evidence for the growth of ions in humid air with low aerosol content has been reported from the naturally radioactive Carlsbad Caverns, New Mexico (Wilkening, 1985), where ion complexes having mobilities considerably smaller than atmospheric small ions have been observed. Decrease in ion mobility results from an increase in ion size.

In the second case listed above, the electric fields caused by the global circuit could affect aerosol charging by modifying the ion environment on the boundaries of clouds where ion concentrations are profoundly asymmetric and the electric fields are enhanced (Carslaw *et al.*, 2002). The global circuit could therefore act to communicate cosmic ray changes throughout the atmosphere, by changes in the conduction current.

### 3.2.2 Nuclear weapons testing

Large-scale changes in atmospheric ionisation occurred as a result of nuclear weapons tests in the 1950s and early 1960s, and radioactive material was injected into the stratosphere. Surface PG values were dramatically reduced (Pierce, 1972). The surface changes resulted from the deposition of radioactivity, greatly increasing the conductivity and reducing the PG near the ground. The PG was reduced to less than 50V.m$^{-1}$ in Scotland in 1963 (Pierce, 1972). Because of surface contamination at many sites, it is unlikely that globally-valid measurements were obtained at surface stations during that period. At Lerwick, where a surface PG reduction was recorded, an accompanying feature was an increase in overcast days, found from examination of the routine solar radiation measurements (Harrison, 2002c). An anomaly in the solar diffuse and direct measurements was also found at Hohenpeißenberg Observatory in Germany by Winkler *et al.* (1998). (Both anomalies occurred before the 1963 eruption of Mt Agung.) If either anomaly was caused by the extra radioactivity in the atmosphere, it would arguably have been more likely to be caused by radioactive ionisation, rather than the supply of aerosol particles from the explosions. This is because the later tests were detonated well above the surface, to reduce the radioactive fallout generated, and therefore produced little aerosol. For the large (megatonne) nuclear explosions occurring at that time, the radioactivity reached the stratosphere, which provided a steady supply of ionisation to the troposphere. With further analysis, data from this period may provide additional evidence for tropospheric particle formation from ions.

### 3.2.3 Atmospheric electricity and global temperature

To consider the relationship between the DC atmospheric electrical circuit and global temperature, either a long series of ionospheric potential measurements or a reconstruction of the global circuit as indicated in Section 3.1 is required. In the absence of either, existing records can therefore be only partially analysed. Section 3.1.1 indicated a basis on which global measurements could be selected. Figure 11(a) shows PG variations found on the basis of diurnal cycle selection, followed by comparison of the mean values resulting with variations at other sites. It shows the average December PG at three European measurement sites: Lerwick (Shetland), Eskdalemuir (Scotland) and Mt Wank (Bavaria) during the 1970s. December is common to both the UK sites as a month having a large positive correlation with the Carnegie curve, and the values used in computing the averages for Mt Wank were selected from days on which the hourly correlation with the Carnegie curve was greater than 0.8. There is a common minimum to all three sites in December 1978, which supports the data selection approach. Additionally, Lerwick and Mt Wank show a short decline from 1980 followed by a rise in 1983 (Harrison, 2004b).

Figure 11 (b) shows the variation in monthly average temperatures for the same Decembers as in Figure 11 (a), using global temperature, northern hemisphere temperatures and southern hemisphere temperatures from land-based meteorological stations (Peterson and Vose, 1997). Although this is only a short period of data, there is a positive correlation between the PG and southern hemisphere temperature for the two UK sites, particularly Lerwick. This is physically reasonable, as the southern hemisphere lightning approaches its seasonal maximum in December, contributing a greater lightning flash rate than the northern hemisphere (Figure 8a ).

### 3.2.4  Effects of global temperature on lightning and the global circuit

A sharply non-linear relationship has been found between lightning flash rate and (wet bulb) temperature (Williams, 1994, 1999), related to the suggestion that local and global temperatures could be monitored sensitively by monitoring lightning and the global circuit (Williams, 1992; Price, 1993). Schumann resonance observations also show a strong correlation with upper troposphere water vapour (Price, 2000), further emphasising the surface temperature-dependence of thunderstorms, and, possibly, electrified shower clouds.

Price (1993) predicted that a 1% (~3K) increase in global surface temperatures could result in a 20% (~50kV) increase in ionospheric potential. Subsequent work by Markson and Price (1999) also found a positive correlation between $V_I$ and global temperature, as derived by satellites. Figure 11 ( c) accordingly shows annual temperatures and the annual average ionospheric potential when available (Markson, 1985). There is qualitative agreement between the Lerwick December average PG and the annual $V_I$ (Harrison, 2004a) indicating that the values are showing the global circuit variations, but a relationship between $V_I$ and annual average temperature anomalies is not so apparent. In part, this may be due to the different importance of different regions on the electrified shower clouds and thunderstorms not captured by the temperature measurements, or, more probably, the poor sampling of $V_I$ temporally. Corrections to the $V_I$ measurements, to allow for differences in the diurnal temperature cycles and the diurnal charge production cycles, have been proposed. From these it has been concluded that $V_I$ is controlled by continental temperatures (Markson, 2003).

### *3.3  Cosmic rays and cloud variations*

A remaining source of electrical changes in the atmosphere arises from the modulation in ion production by cosmic rays (see also Section 3.1.2). Following the early suggestion of Ney (1959), studies have sought to link changes in cosmic rays with atmospheric parameters, such as clouds. The cosmic ray data from the University of Chicago's neutron counter at Climax and other stations are widely available.

### 3.3.1  Surface cloud observations and transient cosmic ray changes

Pudovkin and Veretenko (1995) reported high latitude surface cloudiness observations. They found a short-term decrease in high cloud (cirrus) arising from short periods of reduced cosmic rays ("Forbush Decreases" which are due to increases in the solar wind) using superposed epoch analysis. The work assumed that the cloud observations were weighted towards high clouds.

### 3.3.2  Satellite observations and solar cycle cosmic ray changes

As indicated in Section 1.1, area-averaging is likely to be necessary to reduce the natural variability in the atmosphere sufficiently for atmospheric electrical effects to emerge. An analysis by Marsh and Svensmark (2000), shows that cosmic rays and low (*i.e.* with altitudes less than about 3 km) clouds observed by global satellites are strongly correlated around the cosmic ray minimum of 1990. The correlation was originally presented in 1997 (Svensmark and Friis-Christensen, 1997) and the limitations of the data analysis and acquisition have been the subject of considerable discussion (Kernthaler *et al.*, 1999; Kristjánsson and Kristiansen, 2000). The cloud data used was the D2 dataset of the International Satellite Cloud Climatology Project (ISCCP) (Rossow and Schiffer, 1991), which uses algorithms to identify high cloud (at heights with equivalent pressures of 50-440hPa) and low cloud (680-1000hPa). There is no correlation with high clouds (Kristjánsson and Kristiansen, 2000). In the North and South Atlantic areas, where cosmic ray correlations were found, Sun and Bradley (2002) again report the greatest effect in low clouds. A explanation for a differential cosmic ray effect on low and high clouds has been given in terms of ion-induced particle production by Yu (2002). Yu (2002) argues that the recombination rates of ions are proportional to their concentrations, therefore the ion lifetime is greater at lower altitudes, where the ionisation rate from cosmic rays is smaller. More particles suitable as cloud condensation nuclei would therefore be formed in the lower atmosphere.

### 3.3.3  Evidence from past climates

On very much longer timescales, extending back 500 million years, Shaviv and Veizer (2003) have reported a negative correlation between cosmic rays and global temperature, for cosmic ray changes generated by the solar system passing through the spiral arms of the galaxy. A similar mechanism to that of Marsh and Svensmark (2000), ion-induced CCN formation, has been suggested to explain the finding, in which the CCN lead to more cloudiness, reduced insolation and lower temperatures. Long timescale changes in atmospheric composition and the geomagnetic field may, however, invalidate the connection.

## 4  Conclusions

The atmospheric electrical circuit has probably not remained constant over time. Its long-term variations have only recently begun to be considered and present a new research field to atmospheric science (Harrison, 1997c). Associated with this is the need to assess the importance of global electrical changes in providing a physical mechanism for amplifying the small changes associated with solar variability to give larger atmospheric and climate effects. Physically-plausible mechanisms linking cloud processes with the background electrical properties have been identified in recent work (Carslaw *et al.*, 2002), but the theoretical models from which the magnitudes of the effects concerned can be calculated are only at an early stage. Indications in several data sets suggest that variations in cosmic rays, clouds, the atmospheric electric circuit and global temperature, show at points, close relationships. Currently these can be only partially explained and theory is required if the statistical relationships found are to be understood physically. Figure 12 presents a summary of the global circuit and global climate interactions discussed in the paper.

Further progress requires the integration of different areas of geophysics, which, in many cases, are individually relatively well understood. Taken together, however, a new synthesis of cloud, climate and aerosol physics, atmospheric electricity and atmospheric physics is necessary. Areas which appear particularly fruitful to consider are:

- Reconstruction of past changes in the atmospheric electrical circuit on monthly, annual and decadal timescales, for comparison with global parameters such as global temperatures, cosmic rays and cloud amounts.
- Measurements or modelled distributions of the spatial and temporal changes in ionisation (especially from cosmic rays) in the cloud-forming regions of the atmosphere, for comparison of cloud, ion and aerosol properties in the same regions.
- The electrical coupling between non-thunderstorm clouds and the background electric field needs to be investigated at an electromagnetic level, theoretically and experimentally.
- Research into the basic properties of the atmospheric electric circuit, in terms of its time response to transient events in ionisation (*e.g.*, from nuclear weapons tests and sudden cosmic ray decreases).

There is actually a considerable quantity of disparate surface atmospheric electrical data available, together with occasional aircraft and balloon campaigns. A problem is that these measurements are in different archives and libraries, and very few are available digitally. The data need to be combined to facilitate further analysis. Satellite cloud measurements provide high quality measurements of cloud abundance and type. Integrating satellite and atmospheric electrical data is an essential step in quantifying the electrical influences: however, the meteorological variability in the climate system complicates the extraction of small signals, such as those associated with solar variability. This may be especially true if the timescales of ion-induced aerosol and cloud formation are comparable with the typical timescales of atmospheric changes associated with weather systems, as seems likely. Constructing a numerical model of cloud microphysics, including the electrical interactions of aerosols, ions and water droplets in stratiform clouds presents one example of a quantifiable approach. Reconstruction using proxies for past atmospheric electrical changes are likely to prove highly valuable, but extension of the basic measurements of atmospheric electricity to the global scale for comparison is also essential.

Despite the difficulty in identifying cause and effect in a chaotic system such as the atmosphere, it remains possible that the global atmospheric electrical circuit provides a neglected feedback in the climate system, and with it, an amplification of the solar variability signal in the climate records. This is the principal reason why the topic now deserves further exploration.

**Acknowledgements**
Dr Jürg Beer of the Swiss Federal Institute for Environmental Science and Technology kindly provided the ice core Be-10 data on which Figure 10(b) is based. The UK Met Office obtained the long series of atmospheric electrical data at Lerwick, Eskdalemuir and Kew. Figure 8(a) was based on the v0.1 gridded satellite lightning data, produced by the NASA LIS/OTD Science Team (Principal Investigator, Dr. Hugh J. Christian, NASA Marshall Space Flight Center) and provided by the Global Hydrology Resource Center (http://ghrc.msfc.nasa.gov). I am grateful to Alison Sutton (Department of Meteorology Librarian) and Ian McGregor (National Meteorological Archive) for their help in locating sources of early data.

**Figure captions :** *The global atmospheric electrical system and climate*

Figure 1. Overview of the processes considered in this paper, which link the global atmospheric electrical system with climate. Solid arrows indicate flows of energy (thin lines) or charge (thick lines). Dashed arrows indicate suggested influences and solid arrows established physical mechanisms.

Figure 2. Diurnal variation in Potential Gradient measured by Simpson (1906) at Karasjok, Lapland, between 1903 and 1904. The diurnal variation shows a minimum before dawn and a maximum in the early evening. The amplitude of the diurnal cycle is greatest in winter and smallest in summer.

Figure 3. Principles of early apparatus for surface atmospheric electrical measurements. (a) **Potential Gradient** (PG). A well-insulated collector (radioactive probe, passive sensor, flame or water dropper) is positioned at a height $z$ and allowed to acquire the atmospheric potential $V_0$. The potential of the collector is measured, relative to the surface, using an ultra-high impedance electrometer voltmeter. The potential gradient is found by a correction to $V_0/z$. (b) **Air conductivity**. Air conductivity may be found using a cylindrical capacitor or *Gerdien tube*, through which air is drawn past a well-insulated central electrode. A bias voltage is applied between the electrodes, and the current of small ions passes between the electrodes is measured; this is proportional to the air conductivity. (c) **Air-Earth current**. Wilson's apparatus for air-Earth current consisted of a portable gold leaf electrometer, connected to a horizontal, circular collecting plate, surrounded by an earthed guard ring. The apparatus also included a battery and a variable capacitor (or *compensator*), by which the potential of the collecting plate could be changed, and a brass cover for the collecting plate. The rate of change in plate potential was used to determine the air-Earth current.

Figure 4. (a) Standard diurnal variation in Potential Gradient (Carnegie curve) and global thunderstorm area (from Israel, 1973). (b) Carnegie curve Potential Gradient (expressed as % of mean value in (a) ) plotted as a function of thunderstorm area.

Figure 5 (a) Relative diurnal variation in Potential Gradient found by averaging hourly values from all fair weather days 1968-1971 at Lerwick, Shetland and the ionospheric potential as measured by Mühleisen and Fischer at Weissenau, Germany, with 25 balloon ascents between 13 March 1968 and 24 June 1971. (Note that the time plotted is the *launch* time of the balloon sounding: this causes an offset in the cycle to appear.) The standard Carnegie variation is also shown. (b) Time series of six-hourly average values of PG obtained at Eskdalemuir and on the *Carnegie* (in the western Atlantic), during September 1928. PG measurements from Lerwick for the same period are shown, and the hourly values from *Carnegie* on 15 September 1928 plotted. *Reprinted from Harrison (2004a), with permission from Elsevier.*

Figure 6. Simultaneous diurnal variation of potential in summer 1893, at the surface in Central Paris, and from a water dropper instrument at the top of the Eiffel Tower *Reprinted from Harrison and Aplin (2003), with permission from Elsevier.*

Figure 7. Ionospheric potential measured by Mühleisen and Fischer using synchronous balloon soundings on the *Meteor* (Atlantic Ocean) and at Weissenau (South Germany)

(data from Budyko, 1971). The PG at Lerwick, when available for synchronous hours with no hydrometeors, is also shown.

Figure 8. (a) Seasonal variation in total lightning flash rate obtained by smoothing (using a 55 day moving average to correct for the satellite orbital parameters) the annual climatology from NASA's Lightning Imaging Sensor (LIS) and Optical Transient Detector (OTD). (b) Relative Carnegie diurnal variation as for DJF (December-January-Feburary), MAM (March-April-May), JJA (June-July-August) and SON (September-October-November). (Data from Torreson, 1946, with hourly values > 250V/m excluded.)

Figure 9. Summary of periods for which long-term surface data (mostly measurements of the Potential Gradient) are available, during the nineteenth and twentieth centuries.

Figure 10. (a) Potential Gradient measurements (expressed as % of the long-term mean for the months selected) from the three UK Observatories of Kew (June), Eskdalemuir (annual) and Lerwick (June-July-August) during the first half of the twentieth century. (b) Comparison between the Kew (June) and Eskdalemuir (annual) PG changes shown in (a) with the observed concentration of the cosmogenic isotope beryllium-10, as a % of the 1900-1950 mean.

Figure 11. (a) Annual variations in the December average PG measured at Lerwick and Eskdalemuir (Harrison, 2004a) and Mt Wank in the Bavarian Alps (Harrison, 2004b). (b) Global and hemispheric temperature anomalies (from Global Historical Climatology Network GCHN 1701-12/2001 for land-based surface stations, base period 1951-1980), also for December. (c) Annual global and hemispheric temperature anomalies (from GHCN 1701-12/2001 for land-based surface stations, base period 1951-1980) compared with annual averages of ionospheric potential measurements (Markson, 1985).

Figure 12. Summary of the processes linking the global atmospheric electrical circuit with global climate. Thick lines indicate established links and thin lines indicate suggested links.

**Figure 1**

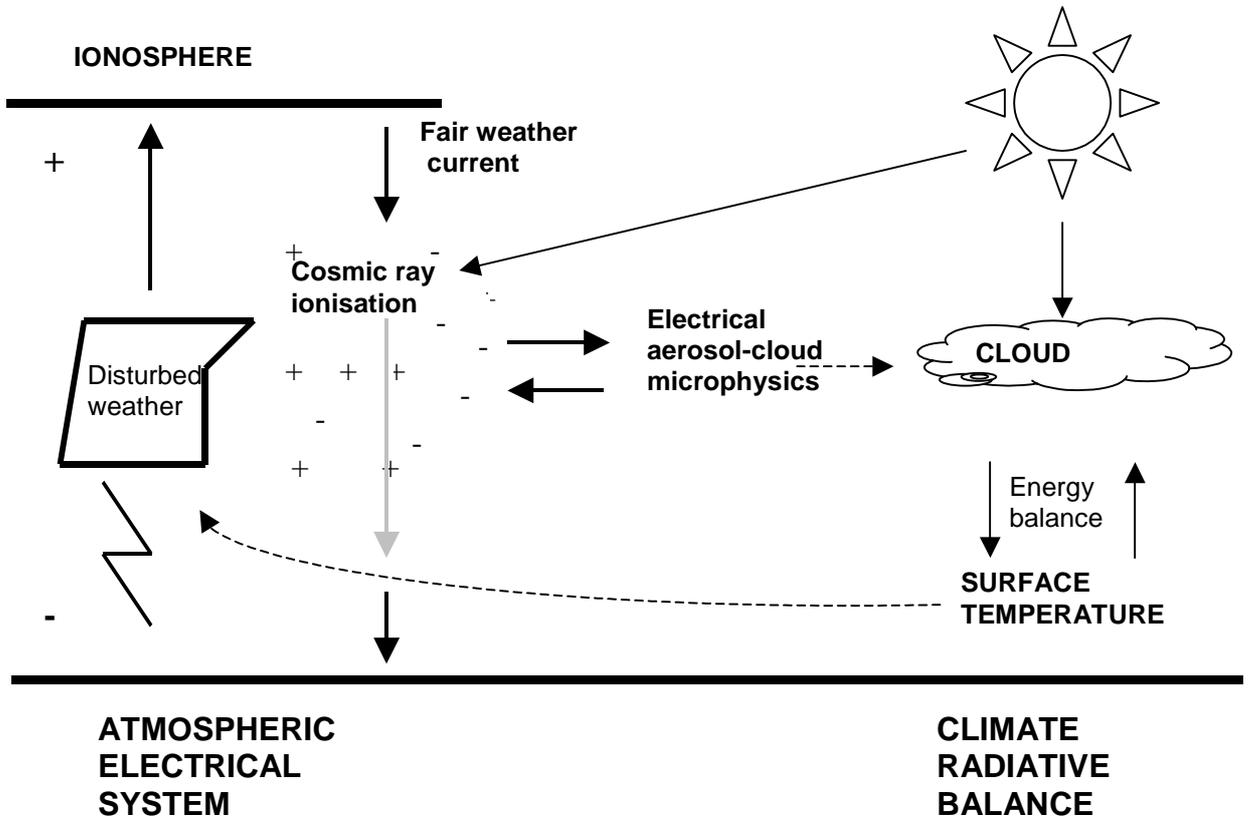

**Figure 2**

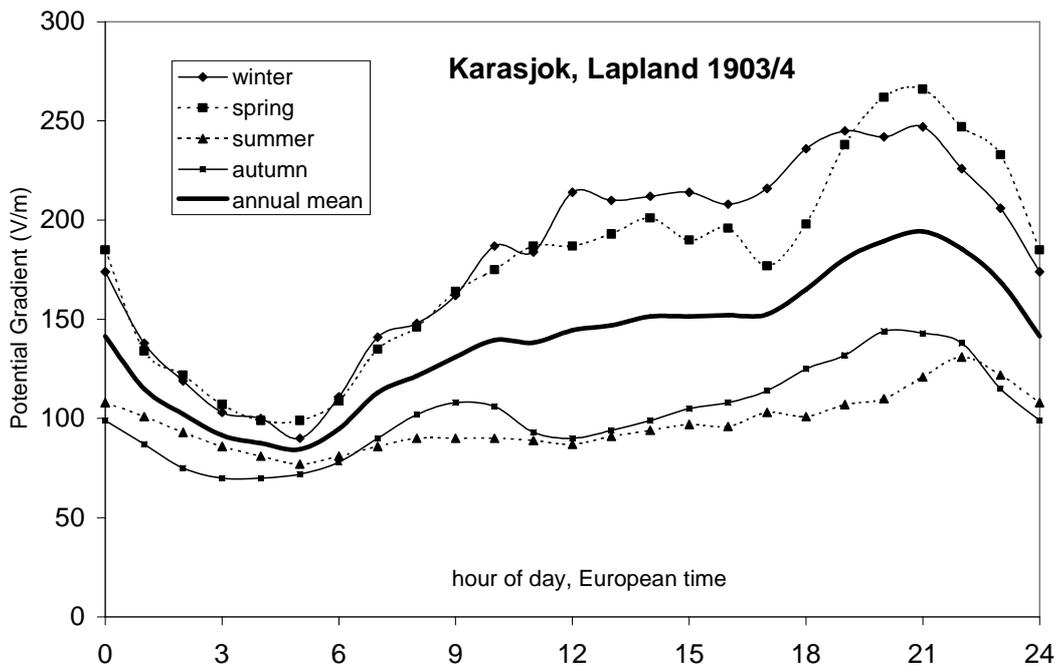

**Figure 3**

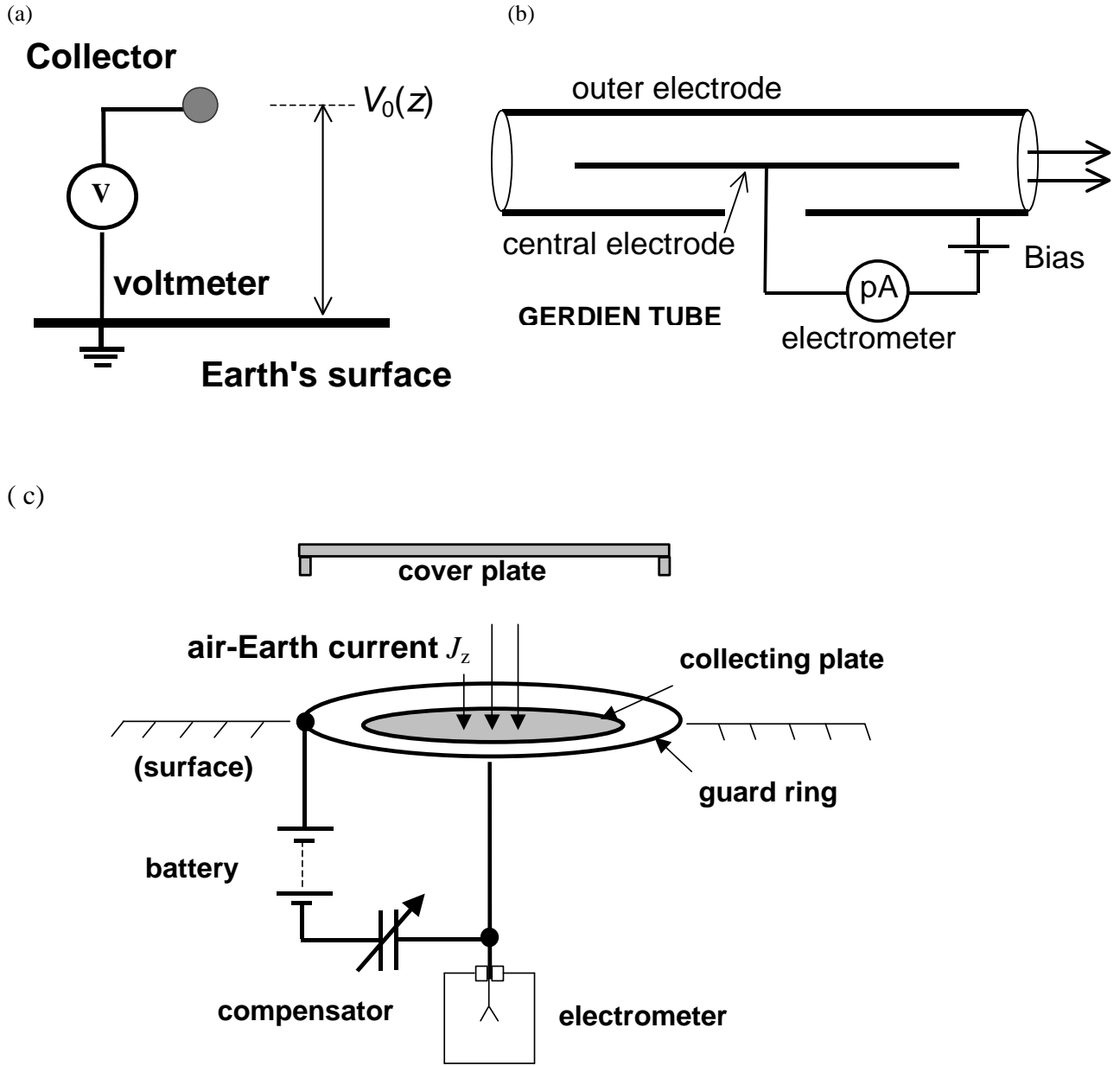

**Figure 4(a)**

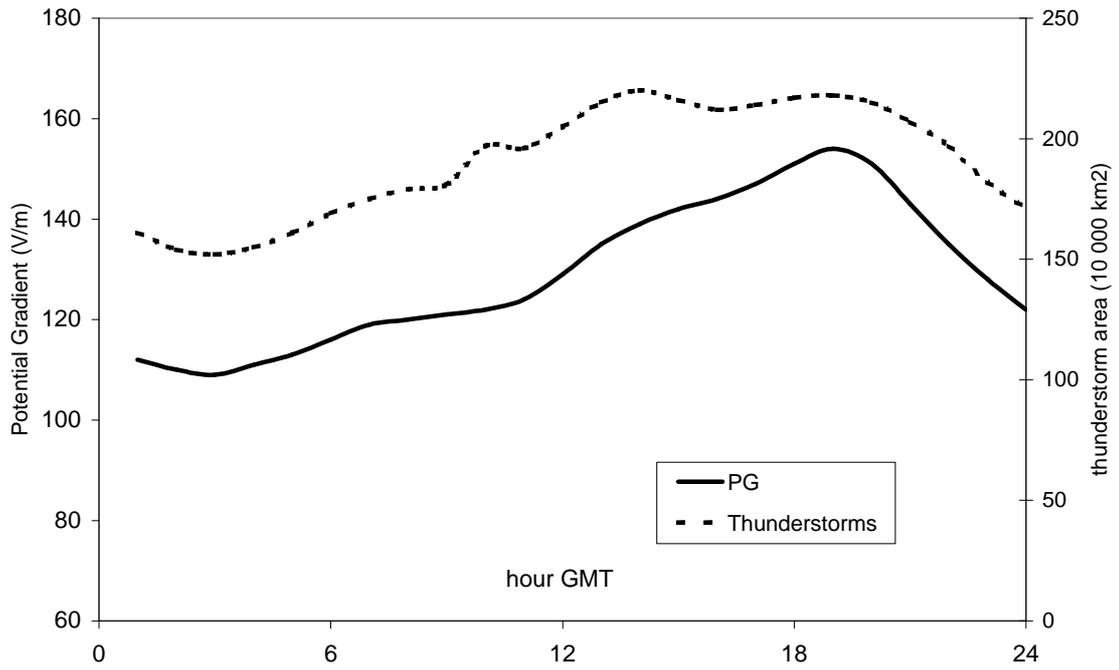

**Figure 4(b)**

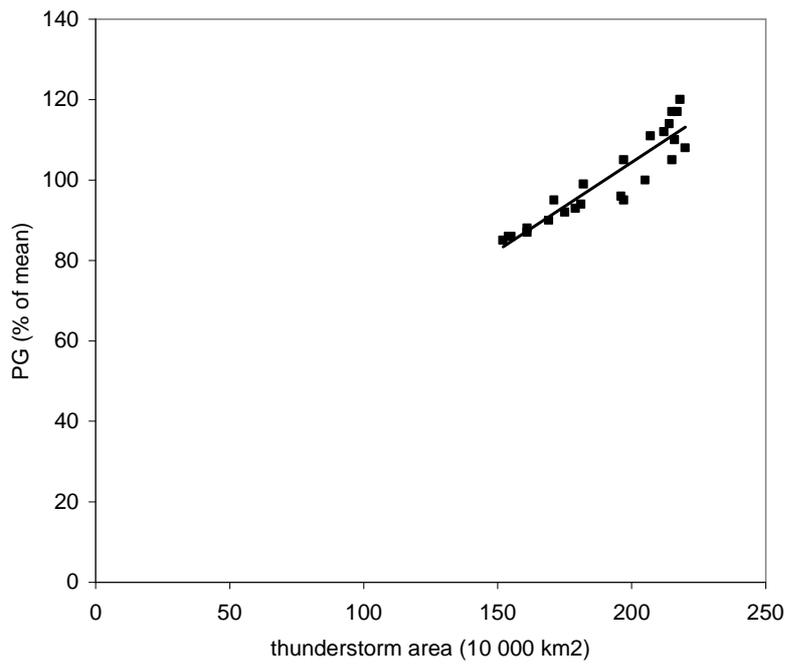

**Figure 5(a)**

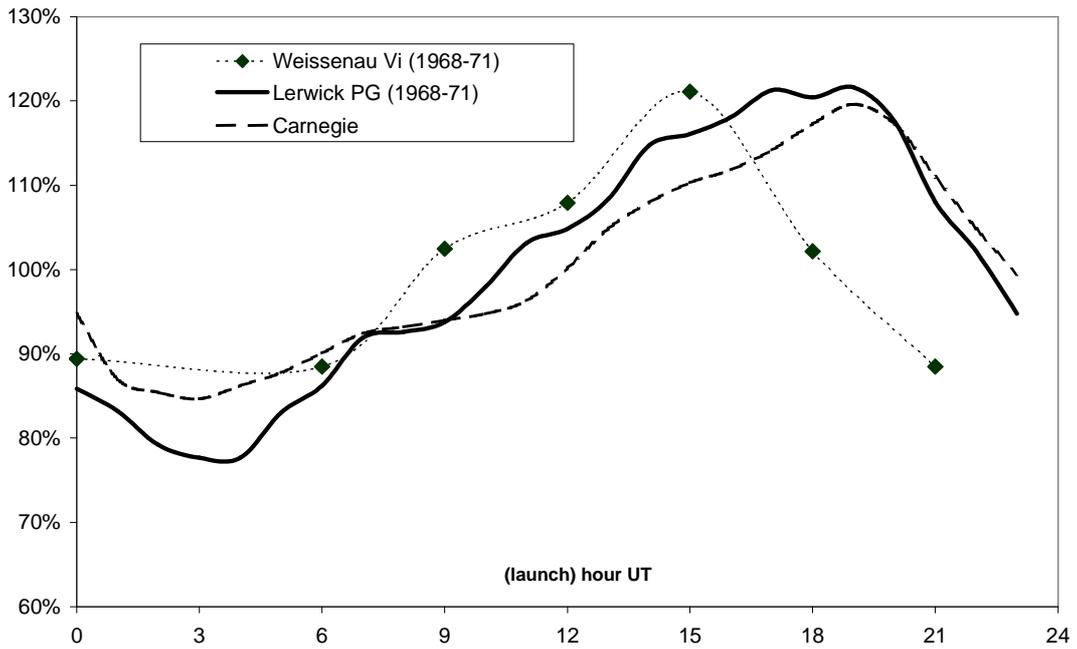

**Figure 5(b).**

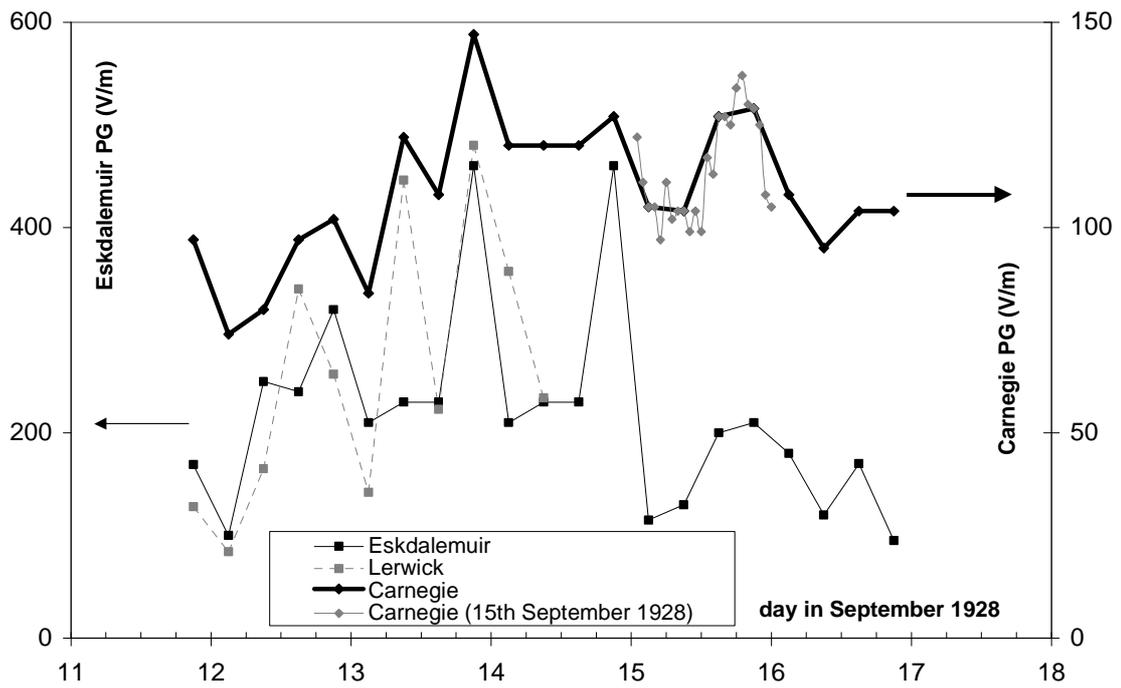

**Figure 6**

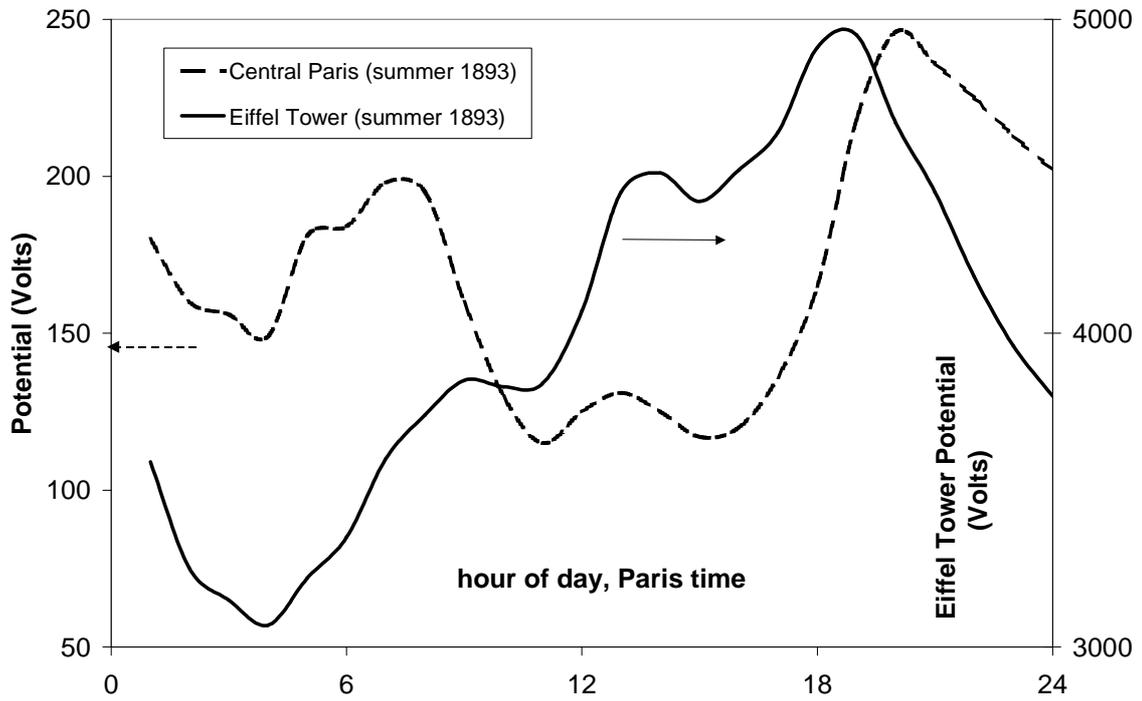

**Figure 7**

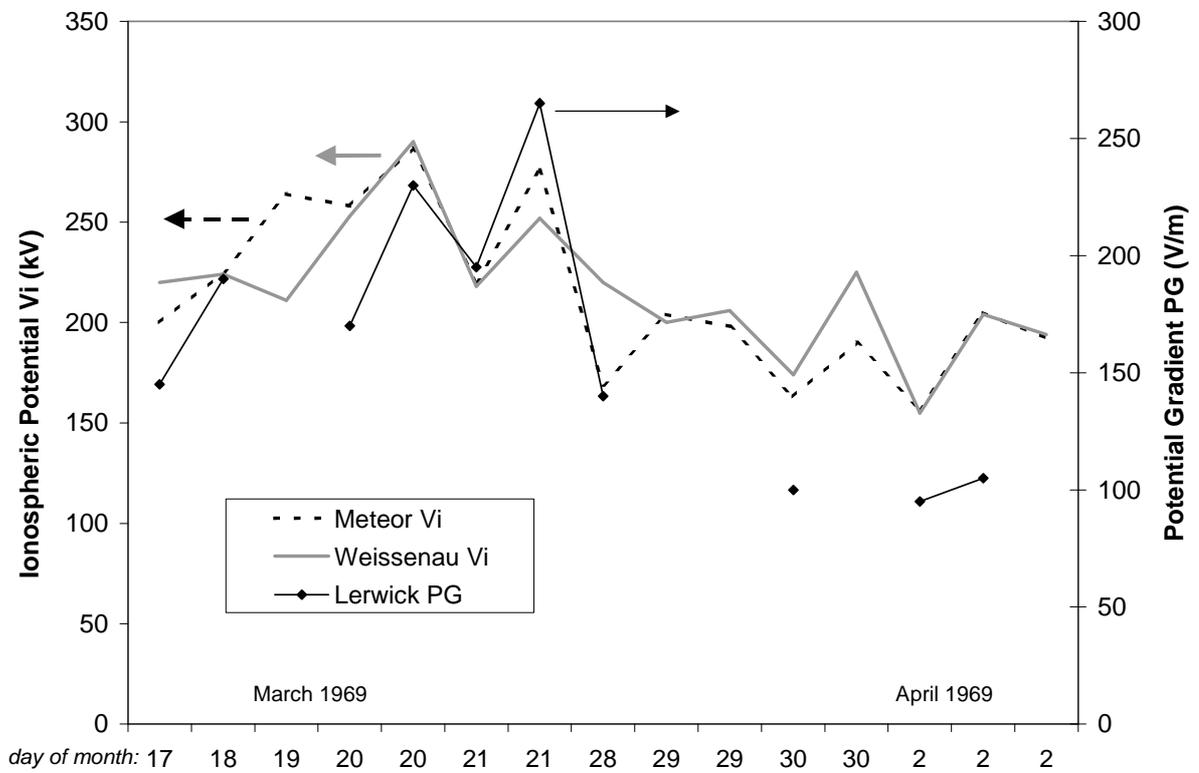

**Figure 8 (a)**

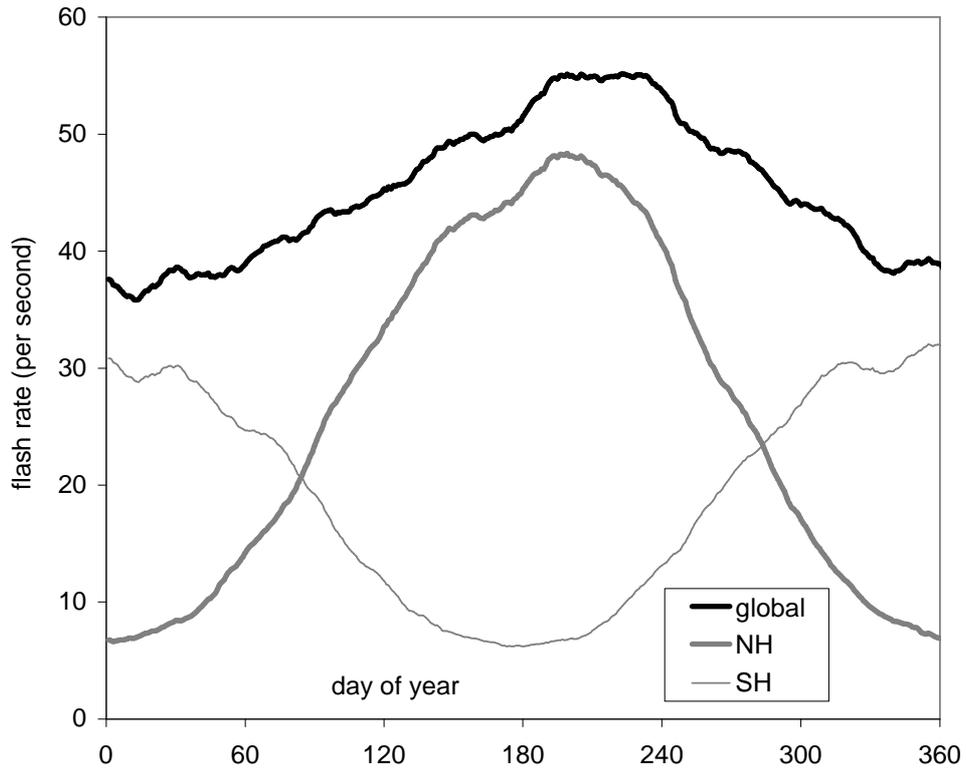

**Figure 8 (b)**

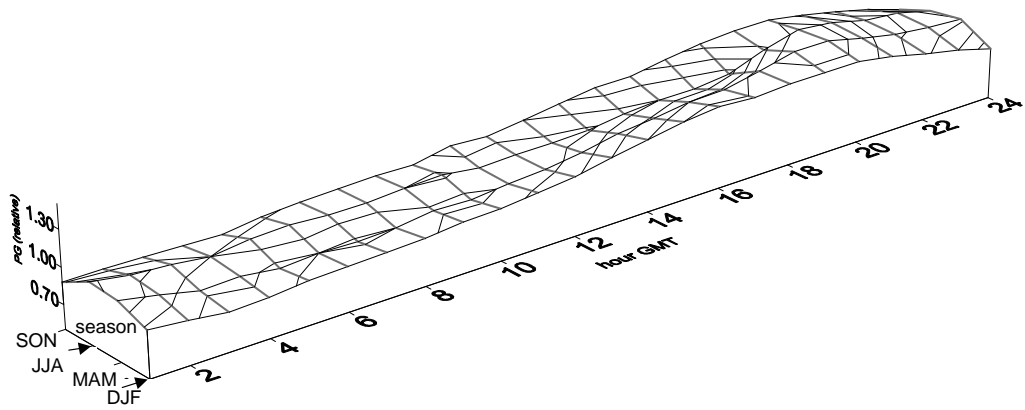

**Figure 9**

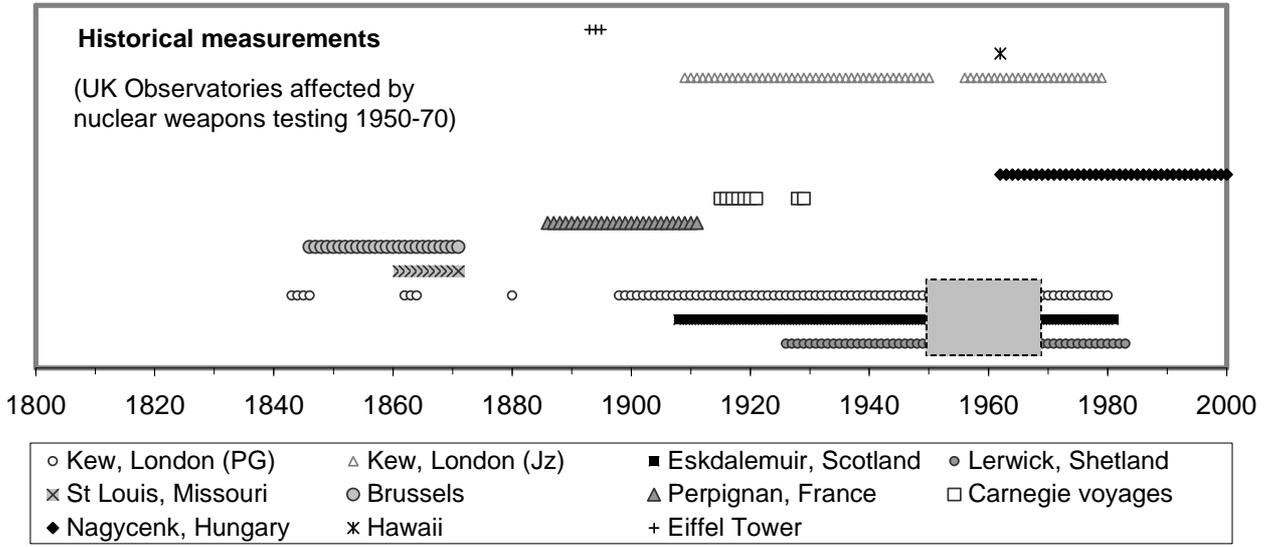

**Figure 10 (a)**

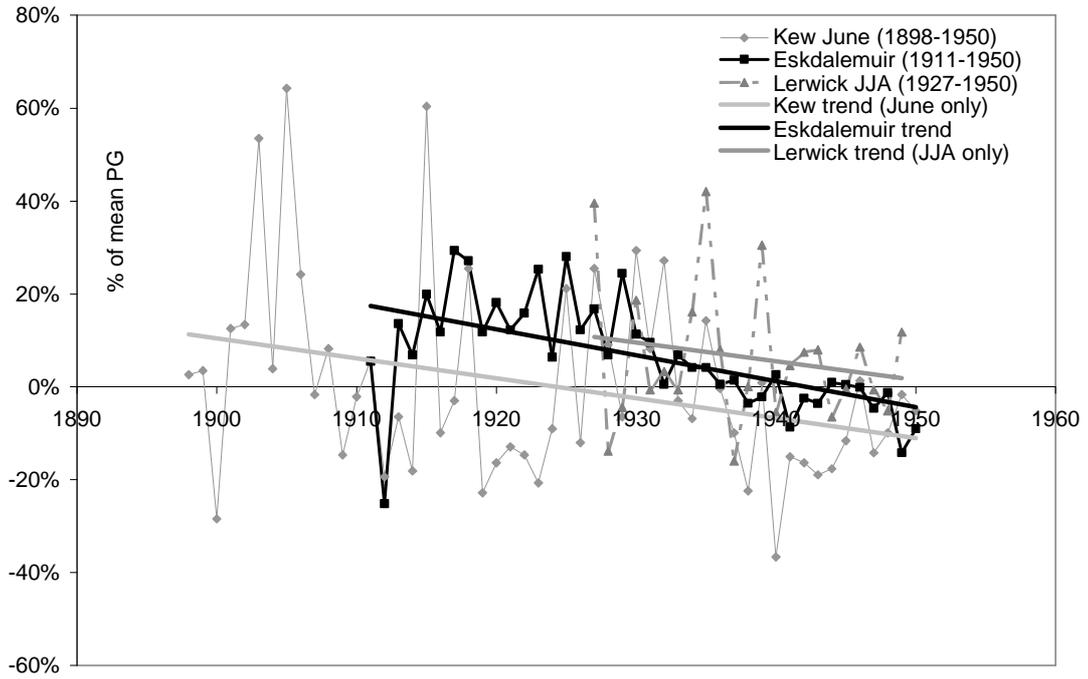

**Figure 10 (b)**

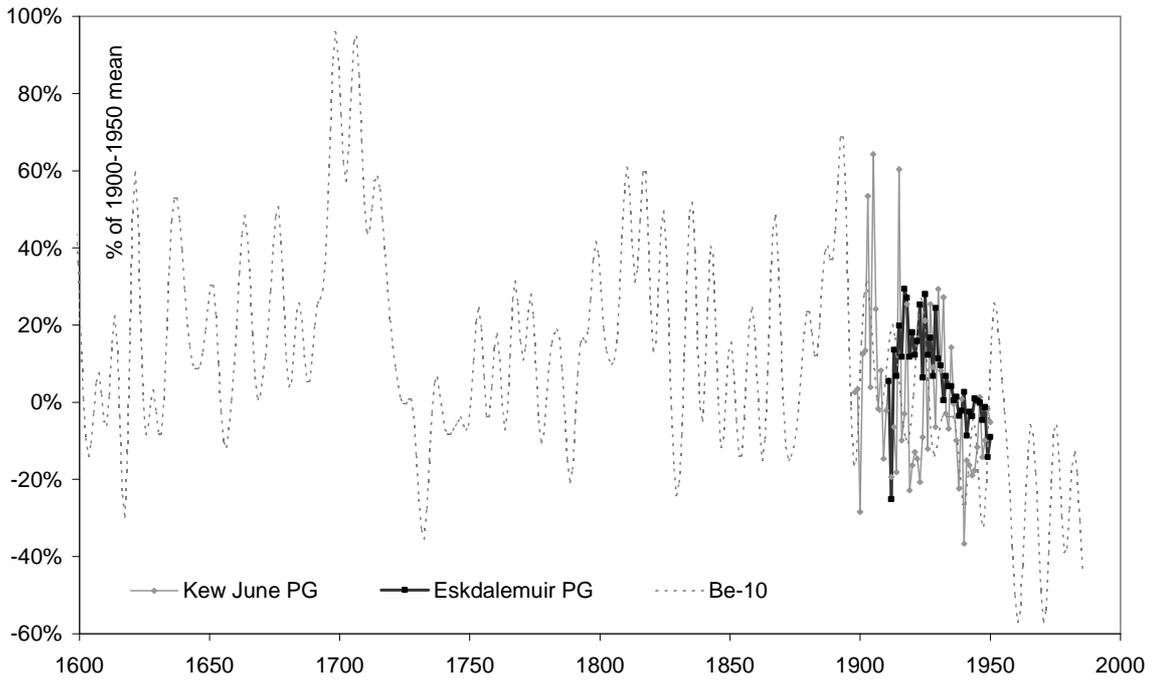

Figure 11 (a)

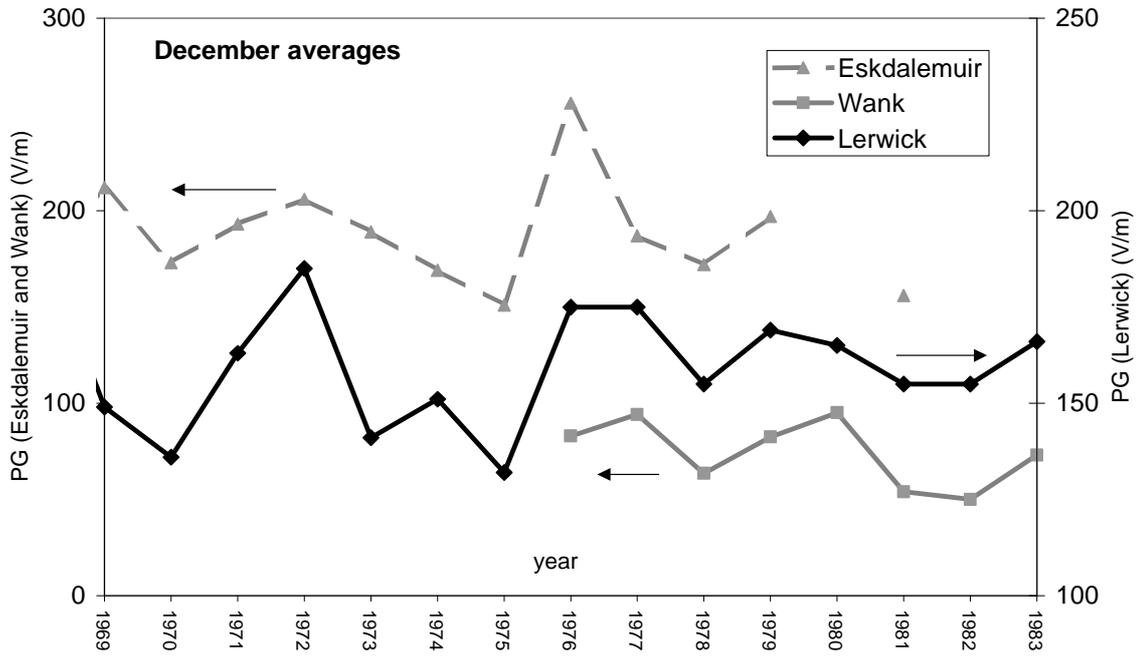

Figure 11(b)

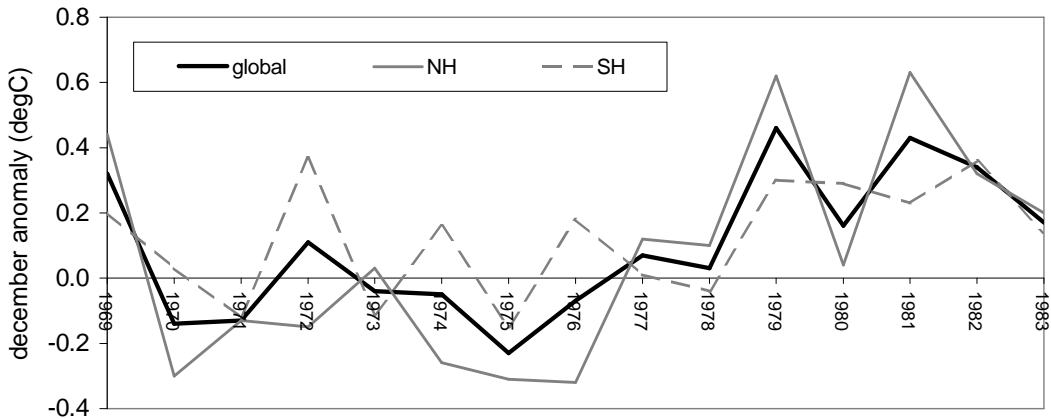

Figure 11(c)

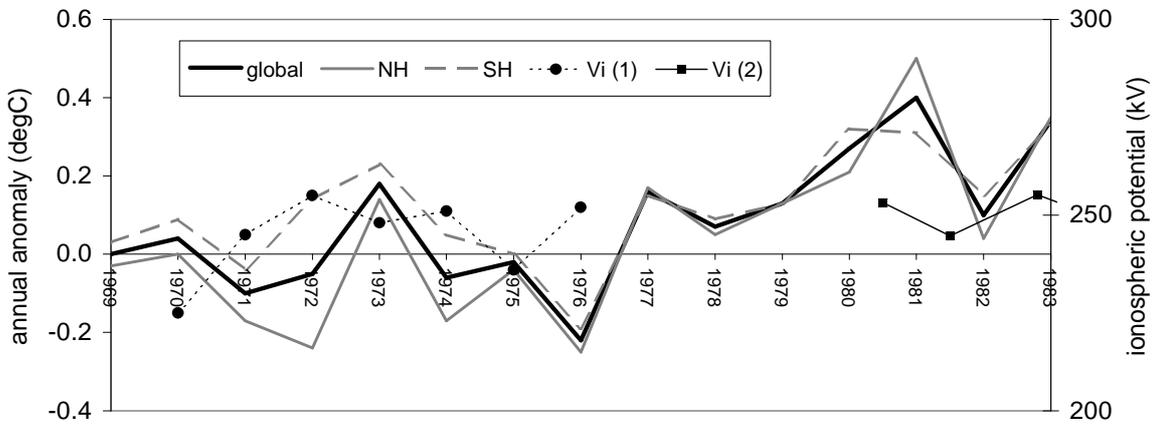

**Figure 12**

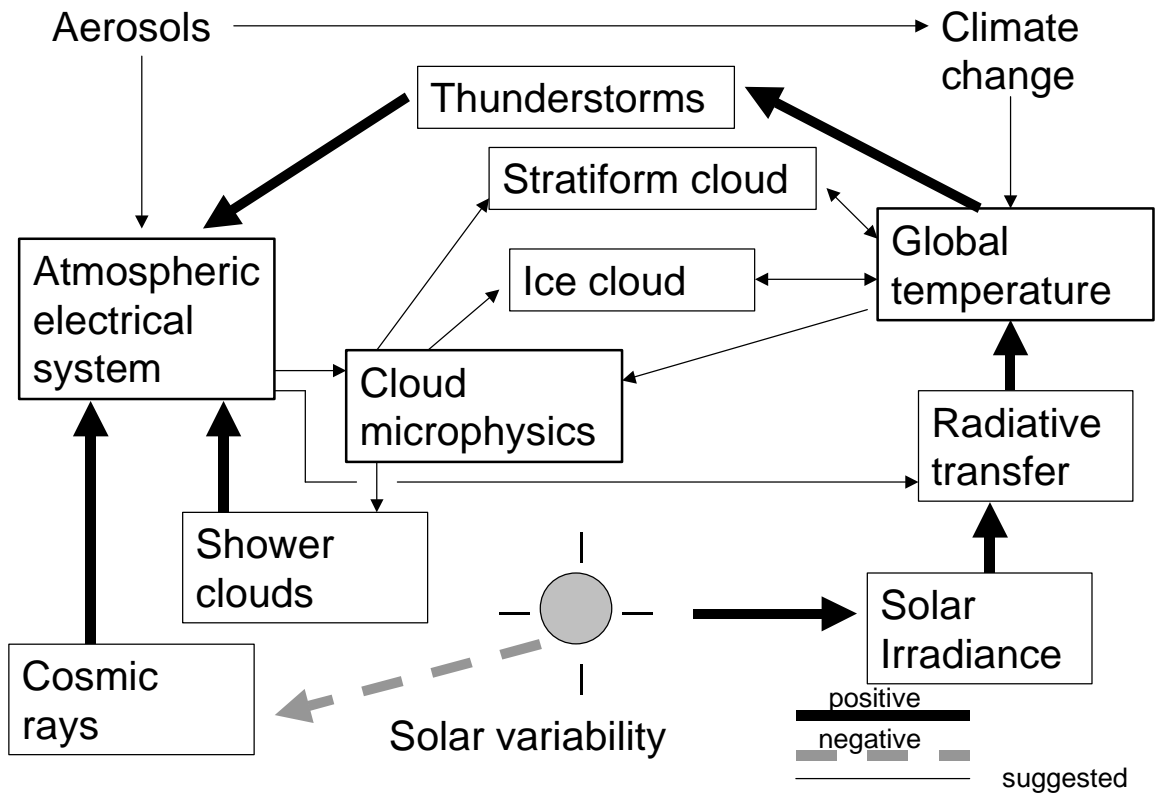

**Footnotes**

---

[1] *i.e.* fair weather

[2] An early photograph survives (Hackmann, 1994)

[3] The PG is defined as $+\mathrm{d}V/\mathrm{d}z$, where $V$ is the potential with respect to earth (at which $V = 0$) at a (positive) height z above the surface. The PG is positive above the surface in fair weather conditions: it is frequently negative in precipitation.

[4] A *thunderday* is a calendar day on which thunder is heard at a meteorological observing station.